%% file: main.tex
\PassOptionsToPackage{hyphens}{url}
\documentclass[sigconf,nonacm]{acmart}
\usepackage{booktabs}
\usepackage{xcolor}
\usepackage{multirow}
\usepackage{amsmath}
\usepackage{graphicx}
\usepackage{microtype}
\microtypesetup{expansion=false}  %
\usepackage{threeparttable}
\usepackage{tabularx}
\usepackage{placeins}  %
\usepackage{balance}
\usepackage{enumitem}
\usepackage{tikz}
\usetikzlibrary{positioning,arrows.meta,calc,fit,backgrounds}
\usepackage[most]{tcolorbox}

\newtcolorbox{rqbox}[1][]{
  colback=gray!7,
  colframe=gray!55!black,
  fonttitle=\bfseries\small,
  title=#1,
  boxrule=0.4pt,
  arc=1pt,
  fontupper=\small,
  left=6pt,right=6pt,top=3pt,bottom=3pt,
  before skip=6pt, after skip=6pt,
  breakable,
  enhanced,
}

\setcopyright{none}
\renewcommand\footnotetextcopyrightpermission[1]{}

\begin{document}

\title{Same Signal, Different Semantics: A Cross-Framework Behavioral Analysis of Software Engineering Agents}

\author{Wei Ma}
\affiliation{%
  \institution{Blekinge Tekniska Högskola}
  \city{Karlskrona}
  \country{Sweden}}

\author{Zhi Chen}
\affiliation{%
  \institution{Singapore Management University}
  \city{Singapore}
  \country{Singapore}}

\author{Jingxu Gu}
\affiliation{%
  \institution{Nanjing University}
  \city{Nanjing}
  \country{China}}

\author{Tianling Li}
\affiliation{%
	\institution{Nanyang Technological University}
	\city{Singapore}
	\country{Singapore}}

\author{Shangqing Liu}
\affiliation{%
  \institution{Nanjing University}
  \city{Nanjing}
  \country{China}}

\author{Lingxiao Jiang}
\affiliation{%
  \institution{Singapore Management University}
  \city{Singapore}
  \country{Singapore}}

\input{sections/abstract}


\maketitle

\input{sections/introduction}

\input{sections/background}

\input{sections/design}

\input{sections/evaluation}

\input{sections/results}

\input{sections/discussion}

\input{sections/threats}

\input{sections/conclusion}

\balance
\bibliographystyle{ACM-Reference-Format}
\bibliography{references}

\end{document}

%% file: sections/abstract.tex
\begin{abstract}
Behavioral studies of LLM-based software engineering agents extract operational rules about which trajectory shapes correlate with higher resolution rates: that a test step follows a code modification, that execution-error cascades are short, or that overall trajectories are compact. Each such rule is typically derived from a single agent framework or a small set of closely related ones, and whether it transfers, in sign as well as magnitude, to structurally different agent designs has not been directly tested.

We address this at ecosystem scale: 64{,}380 SWE-bench runs from 126 agent configurations spanning 43 frameworks, where each configuration pairs an LLM with a framework (e.g., SWE-Agent, OpenHands) that supplies its tools and workflow. We separate framework effects from LLM effects by holding each layer fixed in turn, then test whether behavioral findings transfer across the ecosystem by measuring one behavior--outcome effect per configuration and examining how those effects agree or disagree.

Swapping the framework while the LLM is held fixed produces large behavioral differences in every action feature we measure. But on most behavioral signals, configurations disagree not merely in magnitude but in direction. Error rate is the cleanest case: 47 configurations resolve more issues when their error rate is lower, while 48 resolve more when it is higher. Five other continuous features and three of seven binary behavioral patterns from prior SE literature show similar directional disagreement across configurations. Framework identity, where it can be cleanly attributed, accounts for more of this cross-configuration variation than LLM family: for mean turns, framework explains $64\%$ of the between-configuration variance against the LLM's $10\%$.
The central implication is that the same observable behavioral signal can carry opposite meaning for different agent configurations. Behavioral findings from any single framework therefore warrant cross-configuration validation before being claimed as general; a rule applied without checking framework fit can mislead.

\end{abstract}

\keywords{Software Engineering Agents, Behavioral Analysis, Cross-Framework Study, SWE-bench, Large Language Models}

%% file: sections/introduction.tex
\section{Introduction}
\label{sec:introduction}

LLM-based software engineering agents now resolve over 70\% of real-world GitHub issues on SWE-bench Verified~\cite{jimenez2024swebench}. Performance gains have focused attention on understanding \textit{how} agents behave: multiple studies analyze trajectories to identify success patterns, failure modes, and actionable rules for framework designers~\cite{chen2026process,shepherd2025,agentracer2025,bouzenia2025trajectory,majgaonkar2026trajectory,swebenchpro2025}.

These behavioral studies leave an important external validity question open. Prior work either targets a single framework~\cite{shepherd2025,chen2026process,bouzenia2025trajectory,majgaonkar2026trajectory} or compares a handful of structurally similar ones; neither approach directly measures whether a rule extracted from one configuration survives on a structurally different one. \textit{\textbf{First}}, cross-configuration transferability of findings has not been directly tested. Whether the test-after-modify rule, an operationalization of SHEPHERD's \emph{False Termination} anti-pattern identified on OpenHands~\cite{openhands2024,shepherd2025}, survives on agents with different workflows is a question the community has not answered quantitatively. \textit{\textbf{Second}}, the framework layer of the agent stack and the LLM layer are routinely conflated in such studies, even though they can be separated by design by holding one fixed while the other varies. Without that separation, a finding labeled ``framework matters'' cannot be distinguished from a finding that some specific model--framework combination matters. \textit{\textbf{Third}}, the lack of cross-configuration evidence forces practitioners into a blind design-strategy choice: \textit{when an agent underperforms, should they invest in a stronger LLM or redesign the framework?} These strategies are not equivalent, and the appropriate choice depends on where the heterogeneity lives in the stack; the community has no principled empirical basis for making it.

The same observable run property can predict success for one agent configuration and failure for another, and the layer of the stack that drives that divergence is primarily the framework rather than the LLM. Across our 126 configurations, error rate is the cleanest case: among the configurations with measurable effects, 47 show lower error rate associated with resolution and 48 show the opposite. Five other continuous features and three of seven binary patterns from prior SE literature show the same kind of direction-divided effect. A per-configuration meta-regression then attributes the heterogeneity to the framework where attribution is possible: for mean turns, framework identity explains 64\% of the cross-configuration variance against the LLM's 10\%, the strongest case in a broader pattern of framework-driven variation in trajectory shape and control-flow structure. Behaviorally, the divergence tracks intuitive differences in how different frameworks structure work: an agent with a long-exploratory workflow that generates extended sequences after an error is spiraling through unrecoverable failure, while an agent built around a tight modify-verify loop that does the same has committed to persistent repair. The observable action is identical; the framework determines whether it signals discipline or collapse.

We analyze 64,380 trajectories from 126 $\langle$framework, LLM$\rangle$ configurations across 43 frameworks, exploiting two slices of this matrix that each hold one layer of the stack fixed while letting the other vary. Three \textit{tracer} LLMs (Claude~4 Sonnet, Claude~3.5 Sonnet, and GPT-4o) each appear in six to eight frameworks, so within a tracer the variation we observe must trace to the framework. The complementary slice runs 33 LLMs from 15 model families on a single framework, mini-swe-agent~\cite{minisweagent2024}, where the variation in turn must trace to the LLM. With these two layers separated, we test whether behavioral findings transfer across the 126 configurations through a per-configuration meta-analysis: each configuration contributes one effect size, and we examine the resulting collection of effects in three ways. The magnitude of disagreement is summarized by Higgins' $I^2$~\cite{higgins2003measuring}, the share of between-configuration variation that exceeds sampling noise. The layer responsible for that disagreement is read off two separate meta-regressions, one on framework identity and one on LLM family, yielding the share of cross-configuration variance each layer explains.
Cases where disagreement is not merely large but reverses the sign of the finding are surfaced by the direction split $\langle n_+, n_- \rangle$, the count of configurations whose per-configuration effects are strictly positive versus strictly negative.

Three research questions organize the analysis:

\begin{itemize}
    \item \textbf{RQ1:} What drives behavioral differences across agent configurations: framework design or LLM capability?
    \item \textbf{RQ2:} Which behavioral rules transfer across agent configurations, and which are configuration-specific?
    \item \textbf{RQ3:} Does the predictive value of behavioral features depend on agent configuration, and which layer of the stack drives that dependence?
\end{itemize}

\begin{figure*}[!htb]
\centering
\input{figure1_workflow}
\caption{Research pipeline. Five sequential stages convert raw SWE-bench
trajectories into per-configuration meta-analytic diagnostics that feed the
three research questions. Stage 4 is the only branching point: holding the LLM
fixed (Slice A) isolates framework effects, and holding the framework fixed
(Slice B) isolates LLM effects.}
\label{fig:workflow}
\end{figure*}
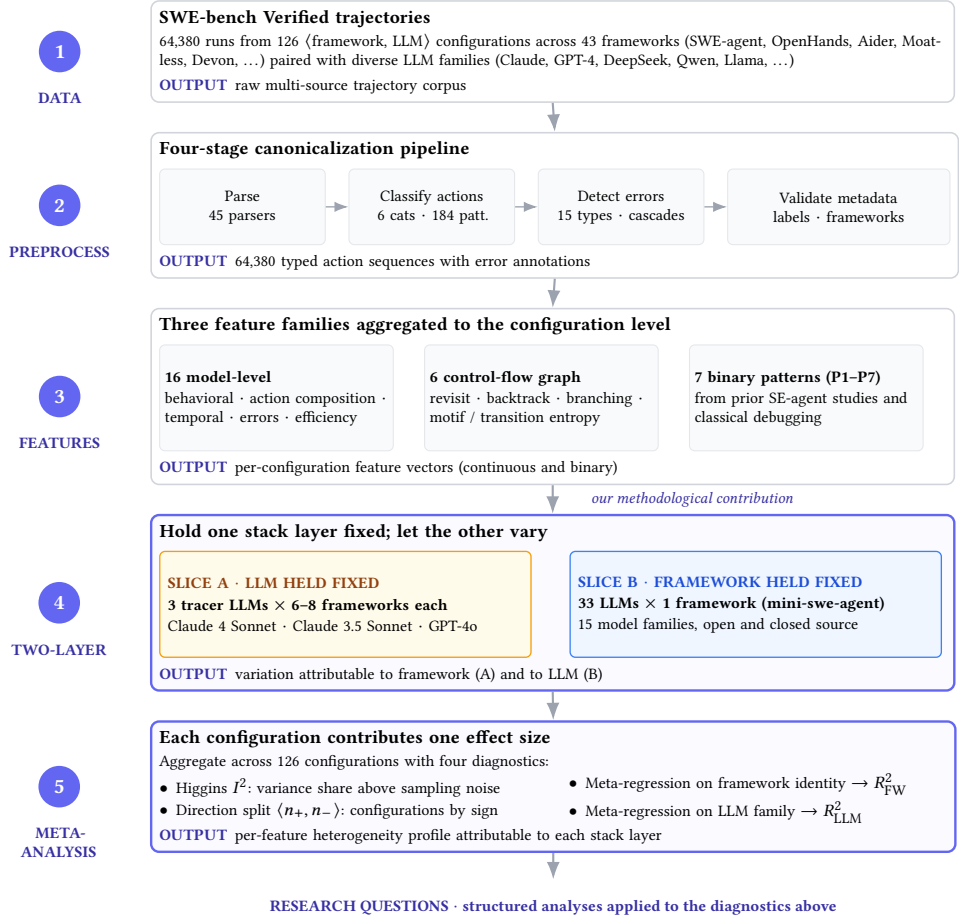

Figure~\ref{fig:workflow} summarizes the research pipeline that links the data, the two-layer design, and the per-configuration meta-analysis to these RQs. RQ1 establishes that framework identity and LLM family are both substantial drivers of agent behavior, motivating the two-layer decomposition. RQ2 applies the per-configuration meta-analysis to seven binary patterns drawn from prior SE literature (debugging, test-driven development, agent efficiency). RQ3 extends it to 13 continuous features and uses meta-regression to attribute heterogeneity to the two stack layers, returning a direct answer to the practitioner's ``upgrade or redesign'' question.

Our contributions are:
\begin{enumerate}
    \item \textit{Behavioral semantics are configuration-conditioned.} We establish that the same observable run property can predict success for some agent configurations and failure for others: across the 126 configurations we study, six continuous features and three of seven binary patterns from prior SE literature show direction-divided effects, including the test-after-modify rule derived from prior single-framework studies. This reframes behavioral analysis from seeking universal rules to calibrating rules against the target framework.

    \item \textit{Cross-configuration validity test at ecosystem scale.} We test rule transferability at ecosystem scale (126 configurations, 43 frameworks) through a per-configuration meta-analysis with framework identity and LLM family as configuration-level covariates. This combination of unit of analysis and scale disentangles the two stack layers in a way single-framework studies cannot.

    \item \textit{Framework as the locus of variation in trajectory shape, with type-conditioned exceptions.} We separately quantify how framework design and LLM capability contribute to cross-configuration variation in the behavior--outcome relationship. Framework identity is the stronger explanatory factor for trajectory-shape features (length, control-flow topology, cascade structure). Across the seven features that clear the permutation diagnostic, framework explains $2.1$ to $6.4$ times the share explained by LLM family, with the strongest gap on mean turns. For action-composition features and raw error counters, neither layer emerges as a clear explanatory factor. Within-type analysis identifies one exception: Type~1 (long-exploratory) trajectories reverse to LLM dominance.

    \item \textit{Framework-conditioned practitioner guidance.} We classify the tested behavioral signals into two transferability classes. Direction-stable signals (shorter trajectories, fewer revisits, lower entropy, lower backtrack rate; $\geq 88\%$ of configurations agree on the sign) carry qualitative principles across frameworks, but per-configuration magnitudes vary so much that numeric targets (e.g., turn-count caps) require per-framework calibration. Direction-unstable signals (error rate, test-after-modify, fast cascade recovery; substantial fractions of configurations show opposite signs) carry no universal rule. This replaces a universal rule book with a framework-aware decision framework.
\end{enumerate}

\paragraph{Paper Organization}
\S\ref{sec:background} reviews related behavioral analyses, positions this work relative to prior single-framework studies, and introduces the $I^2$ heterogeneity tool. \S\ref{sec:design} describes the study design: descriptive trajectory taxonomy, behavioral feature extraction, candidate pattern definitions, and the per-configuration meta-analysis with framework and LLM-family moderators. \S\ref{sec:evaluation} presents our two-dataset design spanning 64,380 trajectories from 43 frameworks. \S\ref{sec:results} presents findings for RQ1--RQ3. \S\ref{sec:discussion} discusses the central findings and provides implications for practitioners, framework designers, and researchers. \S\ref{sec:threats} discusses threats to validity. \S\ref{sec:conclusion} concludes with actionable recommendations.

%% file: figure1_workflow.tex
\begin{tikzpicture}[
  font=\small,
  >=Latex,
]

\definecolor{stagecolor}{HTML}{6366F1}
\definecolor{stagedark}{HTML}{3730A3}
\definecolor{fwcol}{HTML}{F59E0B}
\definecolor{fwbg}{HTML}{FFFBEB}
\definecolor{fwtext}{HTML}{92400E}
\definecolor{llmcol}{HTML}{3B82F6}
\definecolor{llmbg}{HTML}{EFF6FF}
\definecolor{llmtext}{HTML}{1D4ED8}
\definecolor{boxborder}{HTML}{D1D5DB}
\definecolor{subboxbg}{HTML}{F9FAFB}
\definecolor{subboxborder}{HTML}{E5E7EB}
\definecolor{arrowcol}{HTML}{9CA3AF}

\tikzset{
  mbox/.style={
    draw=boxborder, fill=white, line width=0.6pt, rounded corners=3pt,
    inner sep=3pt
  },
  mbox-novel/.style={
    draw=stagecolor, fill=stagecolor!3, line width=0.9pt, rounded corners=3pt,
    inner sep=3pt
  },
  snum/.style={
    circle, fill=stagecolor, text=white, font=\bfseries\small,
    minimum size=5.5mm, inner sep=0
  },
  slabel/.style={
    font=\scriptsize\bfseries, text=stagedark, align=center,
    text width=1.6cm, anchor=north
  },
  shdr/.style={
    font=\bfseries\footnotesize, anchor=north west,
    text width=10.4cm, inner sep=0, align=left
  },
  sbody/.style={
    font=\scriptsize, anchor=north west,
    text width=10.4cm, inner sep=0, align=left
  },
  sout/.style={
    font=\scriptsize, anchor=north west,
    text width=10.4cm, inner sep=0, align=left
  },
  subbox/.style={
    draw=subboxborder, fill=subboxbg, line width=0.4pt, rounded corners=2pt,
    inner sep=2pt, font=\scriptsize, align=center, anchor=north west
  },
  fwbox/.style={
    draw=fwcol, fill=fwbg, line width=0.5pt, rounded corners=2pt,
    inner sep=3pt, font=\scriptsize, align=left, anchor=north west
  },
  llmbox/.style={
    draw=llmcol, fill=llmbg, line width=0.5pt, rounded corners=2pt,
    inner sep=3pt, font=\scriptsize, align=left, anchor=north west
  },
  arr/.style={->, line width=0.8pt, color=arrowcol},
  arrs/.style={->, line width=0.5pt, color=arrowcol},
}

\coordinate (xleft) at (2.0, 0);

\node[shdr] (s1head) at (xleft) {SWE-bench Verified trajectories};
\node[sbody] (s1body) at ($(s1head.south west) + (0,-0.8mm)$)
  {64,380 runs from 126 $\langle$framework, LLM$\rangle$ configurations across
   43 frameworks (SWE-agent, OpenHands, Aider, Moatless, Devon, \ldots) paired
   with diverse LLM families (Claude, GPT-4, DeepSeek, Qwen, Llama, \ldots)};
\node[sout] (s1out) at ($(s1body.south west) + (0,-1.2mm)$)
  {\textcolor{stagedark}{\bfseries OUTPUT}\enspace raw multi-source trajectory corpus};
\begin{scope}[on background layer]
  \node[mbox, fit=(s1head)(s1body)(s1out)] (s1) {};
\end{scope}
\node[snum] at ([xshift=-1.2cm]s1.west) (s1n) {1};
\node[slabel] at ([yshift=-1.5mm]s1n.south) {DATA};

\node[shdr] (s2head) at ($(xleft |- s1.south) + (0,-5mm)$)
  {Four-stage canonicalization pipeline};
\node[subbox, text width=2.05cm, minimum height=10mm] (p1)
  at ($(s2head.south west) + (0,-1.2mm)$)
  {Parse\\\scriptsize 45 parsers};
\node[subbox, text width=2.05cm, minimum height=10mm] (p2)
  at ($(p1.east) + (3mm,0)$) [anchor=west]
  {Classify actions\\\scriptsize 6 cats $\cdot$ 184 patt.};
\node[subbox, text width=2.05cm, minimum height=10mm] (p3)
  at ($(p2.east) + (3mm,0)$) [anchor=west]
  {Detect errors\\\scriptsize 15 types $\cdot$ cascades};
\node[subbox, text width=2.8cm, minimum height=10mm] (p4)
  at ($(p3.east) + (3mm,0)$) [anchor=west]
  {Validate metadata\\\scriptsize labels $\cdot$ frameworks};
\draw[arrs] (p1.east) -- (p2.west);
\draw[arrs] (p2.east) -- (p3.west);
\draw[arrs] (p3.east) -- (p4.west);
\node[sout] (s2out) at ($(p1.south -| s2head.south west) + (0,-1.2mm)$)
  {\textcolor{stagedark}{\bfseries OUTPUT}\enspace 64,380 typed action sequences with error annotations};
\begin{scope}[on background layer]
  \node[mbox, fit=(s2head)(p1)(p4)(s2out)] (s2) {};
\end{scope}
\node[snum] at ([xshift=-1.2cm]s2.west) (s2n) {2};
\node[slabel] at ([yshift=-1.5mm]s2n.south) {PREPROCESS};
\draw[arr] (s1.south) -- (s2.north);

\node[shdr] (s3head) at ($(xleft |- s2.south) + (0,-5mm)$)
  {Three feature families aggregated to the configuration level};
\node[subbox, text width=2.95cm, minimum height=14mm, align=left] (f1)
  at ($(s3head.south west) + (0,-1.2mm)$)
  {\textbf{16 model-level}\\[1pt]\scriptsize behavioral $\cdot$ action composition $\cdot$ temporal $\cdot$ errors $\cdot$ efficiency};
\node[subbox, text width=2.95cm, minimum height=14mm, align=left] (f2)
  at ($(f1.east) + (4mm,0)$) [anchor=west]
  {\textbf{6 control-flow graph}\\[1pt]\scriptsize revisit $\cdot$ backtrack $\cdot$ branching $\cdot$ motif / transition entropy};
\node[subbox, text width=2.95cm, minimum height=14mm, align=left] (f3)
  at ($(f2.east) + (4mm,0)$) [anchor=west]
  {\textbf{7 binary patterns (P1--P7)}\\[1pt]\scriptsize from prior SE-agent studies and classical debugging};
\node[sout] (s3out) at ($(f1.south -| s3head.south west) + (0,-1.2mm)$)
  {\textcolor{stagedark}{\bfseries OUTPUT}\enspace per-configuration feature vectors (continuous and binary)};
\begin{scope}[on background layer]
  \node[mbox, fit=(s3head)(f1)(f3)(s3out)] (s3) {};
\end{scope}
\node[snum] at ([xshift=-1.2cm]s3.west) (s3n) {3};
\node[slabel] at ([yshift=-1.5mm]s3n.south) {FEATURES};
\draw[arr] (s2.south) -- (s3.north);

\node[shdr] (s4head) at ($(xleft |- s3.south) + (0,-5mm)$)
  {Hold one stack layer fixed; let the other vary};
\node[fwbox, text width=4.7cm, minimum height=14mm] (sa)
  at ($(s4head.south west) + (0,-1.2mm)$)
  {{\scriptsize\bfseries\color{fwtext} SLICE A $\cdot$ LLM HELD FIXED}\\[1pt]
   \textbf{3 tracer LLMs $\times$ 6--8 frameworks each}\\[1pt]
   \scriptsize Claude~4 Sonnet $\cdot$ Claude~3.5 Sonnet $\cdot$ GPT-4o};
\node[llmbox, text width=4.7cm, minimum height=14mm] (sb)
  at ($(sa.east) + (5mm,0)$) [anchor=west]
  {{\scriptsize\bfseries\color{llmtext} SLICE B $\cdot$ FRAMEWORK HELD FIXED}\\[1pt]
   \textbf{33 LLMs $\times$ 1 framework (mini-swe-agent)}\\[1pt]
   \scriptsize 15 model families, open and closed source};
\node[sout] (s4out) at ($(sa.south -| s4head.south west) + (0,-1.2mm)$)
  {\textcolor{stagedark}{\bfseries OUTPUT}\enspace variation attributable to framework (A) and to LLM (B)};
\begin{scope}[on background layer]
  \node[mbox-novel, fit=(s4head)(sa)(sb)(s4out)] (s4) {};
\end{scope}
\node[snum] at ([xshift=-1.2cm]s4.west) (s4n) {4};
\node[slabel] at ([yshift=-1.5mm]s4n.south) {TWO-LAYER};
\draw[arr] (s3.south) -- (s4.north);
\node[font=\scriptsize\itshape, text=stagedark, anchor=west, inner sep=0]
  at ($(s3.south)!0.5!(s4.north) + (5mm,0)$)
  {our methodological contribution};

\node[shdr] (s5head) at ($(xleft |- s4.south) + (0,-5mm)$)
  {Each configuration contributes one effect size};
\node[sbody] (s5sub) at ($(s5head.south west) + (0,-0.8mm)$)
  {Aggregate across 126 configurations with four diagnostics:};
\node[font=\scriptsize, anchor=north west, text width=5cm, inner sep=0, align=left]
  (s5l) at ($(s5sub.south west) + (0,-1mm)$)
  {$\bullet$\, Higgins $I^2$: variance share above sampling noise\\[1pt]
   $\bullet$\, Direction split $\langle n_+, n_-\rangle$: configurations by sign};
\node[font=\scriptsize, anchor=north west, text width=5cm, inner sep=0, align=left]
  (s5r) at ($(s5l.east) + (4mm,0)$) [anchor=west]
  {$\bullet$\, Meta-regression on framework identity $\to$ $R^2_{\text{FW}}$\\[1pt]
   $\bullet$\, Meta-regression on LLM family $\to$ $R^2_{\text{LLM}}$};
\node[sout] (s5out) at ($(s5l.south -| s5head.south west) + (0,-1.2mm)$)
  {\textcolor{stagedark}{\bfseries OUTPUT}\enspace per-feature heterogeneity profile attributable to each stack layer};
\begin{scope}[on background layer]
  \node[mbox-novel, fit=(s5head)(s5sub)(s5l)(s5r)(s5out)] (s5) {};
\end{scope}
\node[snum] at ([xshift=-1.2cm]s5.west) (s5n) {5};
\node[slabel] at ([yshift=-1.5mm]s5n.south) {META-ANALYSIS};
\draw[arr] (s4.south) -- (s5.north);

\node[anchor=north, font=\scriptsize\bfseries, text=stagedark]
  (rqlbl) at ($(s5.south) + (0,-5mm)$)
  {RESEARCH QUESTIONS $\cdot$ structured analyses applied to the diagnostics above};
\draw[arr] (s5.south) -- ([yshift=1.5mm]rqlbl.north);

\end{tikzpicture}

%% file: sections/background.tex
\section{Background}
\label{sec:background}

\paragraph{Structural diversity in SE agents.} The SE agent ecosystem spans structurally distinct designs. SWE-Agent~\cite{sweagent2024} and OpenHands~\cite{openhands2024} place the LLM in an open-ended reasoning loop with shell or tool-call interfaces; the agent iterates until it decides to stop. Agentless~\cite{xia2023agentless} sits at the opposite end of the loop-based versus pipeline spectrum, removing the loop entirely: a hierarchical localization step (file $\to$ function $\to$ line) narrows the search space, then a patch is generated and validated in a single pass. Other designs introduce orthogonal variations: AutoCodeRover~\cite{zhang2024autocoderover} interposes program analysis before patching, and MASAI~\cite{masai2024} decomposes the task across sub-agents with distinct roles. These designs do not merely differ in implementation; they impose different structural constraints on how agents can behave, which raises the central question this paper addresses: do behavioral findings transfer across them?

\paragraph{Prior behavioral analyses.} Prior work has produced rich accounts of SE agent behavior, but within narrow architectural scope. SHEPHERD~\cite{shepherd2025} identifies failure patterns that correlate with resolution across 18 LLMs and 3.9K trajectories on OpenHands. Chen et al.~\cite{chen2026process} classify Python execution errors across 8 agents (3.9K trajectories). Bouzenia and Pradel~\cite{bouzenia2025trajectory} identify anti-patterns such as repetitive cycles and missing-verification cycles across 3 agents (120 trajectories). Majgaonkar et al.~\cite{majgaonkar2026trajectory} report that failed trajectories are consistently longer across 3 agents on SWE-Bench Lite and Verified ($\approx$2.4K trajectories). Yin et al.~\cite{yin2025frameworks} compare framework effectiveness across seven frameworks on code-centric tasks. AgenTracer~\cite{agentracer2025} attributes failures to specific sub-agents or turns across six benchmark datasets (over 2{,}000 trajectories) via counterfactual replay. None of these studies directly tests whether its behavioral findings transfer across structurally different agent designs. We address this gap with a per-configuration analysis over 126 configurations that quantifies how much each behavioral signal transfers across the ecosystem, and a two-layer decomposition that attributes the remaining variation to framework versus LLM identity.

\paragraph{Design-level vs.\ trace-level classification.} Existing taxonomies of LLM agents classify them by intended design: their reasoning pattern (e.g., ReAct~\cite{yao2022react}), workflow architecture (e.g., Plan-and-Execute), or exposed capability (e.g., tool use). We take a trace-level view instead: because behavioral analysis operates on runtime trajectories, we group agents by observable trace structure such as action format, loop pattern, and trajectory compactness. For instance, mini-swe-agent produces compact short-step trajectories with weaker models but longer, more deliberate ones with stronger models. Our five trajectory types (\S\ref{par:taxonomy}) are defined at this trace level.

\paragraph{$I^2$ heterogeneity.} We quantify cross-configuration transferability by adapting the $I^2$ heterogeneity statistic~\cite{higgins2003measuring} from meta-analysis. In its original setting, $I^2$ measures how much of the variation across independent studies reflects real differences rather than sampling noise; we apply the same logic at the \textit{configuration level}, treating each of the 126 agent configurations as an independent study. Low $I^2$ implies that an effect transfers stably across configurations; high $I^2$ implies that the effect is configuration-dependent. Calculation, classification thresholds, and the complementary direction split $\langle n_+, n_- \rangle$ are given in \S\ref{sec:per_config_meta}.

%% file: sections/design.tex
\section{Study Design}
\label{sec:design}

Every agent configuration combines two choices: a framework (e.g., SWE-Agent, OpenHands) and an LLM (e.g., GPT-4o, Claude~4 Sonnet). When two configurations differ in behavior, either choice could be the cause. A full factorial design crossing all 43 frameworks with all LLMs would require thousands of runs at substantial compute cost, and is unnecessary: the SWE-bench leaderboard already provides 64,380 trajectories from real-world submissions. However, developers submit their best configurations rather than all possible combinations, so the observed framework--LLM matrix is sparse. We instead construct two comparisons that each hold one factor constant. First, a \textit{single-framework baseline}: 33 LLMs all running on the same framework (mini-swe-agent), so any behavioral difference must come from the model. Second, a \textit{multi-tracer natural experiment}: we select three LLMs (which we call \textit{tracers}) that each appear in 6--8 different frameworks, so any behavioral difference within a tracer must come from the framework.

\subsection{Datasets}

\paragraph{Sources.}
Trajectories come from two public repositories: the SWE-bench Experiments repository\footnote{\url{https://github.com/SWE-bench/experiments}} (leaderboard submissions) and the Docent platform\footnote{\url{https://docent.transluce.org/}} (full conversation transcripts). All trajectories target SWE-bench Verified~\cite{jimenez2024swebench} under the 100\% oracle setting (full test visibility). From these sources we derive two subsets for the analyses that follow. The \textit{bash-only} subset (16,522 trajectories, 33 LLMs on mini-swe-agent) holds the framework constant, isolating LLM-level variation; it serves as the controlled baseline for RQ1 and the pattern-definition reference for RQ2. The \textit{verified} subset (47,858 trajectories, 42 frameworks, 93 configurations) supplies the framework diversity required for the per-configuration meta-analysis in RQ2 and RQ3 and for the multi-tracer natural experiment in RQ1. Together the two subsets cover 64,380 trajectories from 126 configurations across 43 frameworks; full dataset statistics appear in \S\ref{sec:evaluation}.

\paragraph{Preprocessing.}
Each trajectory passes through four stages.
\textit{(1) Parsing}: because no common trajectory format exists across the 43 frameworks, we use 45 agent-specific parsers (grouped into 15 format families) that convert raw JSON, YAML, or plain-text logs into structured $\langle \theta, a, o \rangle$ triples.\footnote{Loop-based frameworks define a turn at the LLM-call level (one $\langle \theta, a, o \rangle$ per call), while non-loop pipelines such as Agentless define it at the pipeline-stage level (localization, patch generation, test generation); we therefore avoid magnitude-based comparisons of absolute turn counts across the two designs.}
\textit{(2) Action classification}: each action is mapped to one of six semantic categories (Table~\ref{tab:action_types}) by matching the command string against 97 tool-call signatures (e.g., \texttt{file\_viewer}, \texttt{code\_search}) and 87 bash command names (e.g., \texttt{grep}, \texttt{sed}).
\textit{(3) Error detection}: error-producing turns are identified by regex over environment responses and categorized into 15 types (test failure, traceback, syntax error, timeout, etc.). A run of consecutive error-producing turns within a trajectory is an \textit{error cascade}; cascade length and cascade rate (defined in Table~\ref{tab:features}) quantify how errors propagate.
\textit{(4) Metadata validation}: resolution status is cross-referenced with SWE-bench leaderboard results; model and framework metadata are extracted from directory structure and trajectory headers.
We validate the pipeline on two axes. For accuracy, the classifier achieves Cohen's $\kappa > 0.85$ on 500 manually annotated turns across 5 frameworks. For coverage, 122 of 132 configurations have an unknown-action rate of at most $5\%$ (median 0.4\%); the one outlier exceeding $50\%$ (a tool-call agent at $65\%$ unknown) is excluded from action-distribution analyses but retained for resolution-rate comparisons.

\paragraph{Inclusion criteria.}
The 126 configurations used in our analysis were not the entire universe of public SWE-bench submissions; they result from three sequential decisions made during data collection, each tied to an operational step in our pipeline. (1)~\textit{Benchmark and setting alignment}: we retain only submissions targeting SWE-bench Verified under the 100\% oracle setting, excluding Lite, Full, and other oracle configurations so that difficulty and information availability are held constant. (2)~\textit{Trajectory parseability}: a submission is included only if our parser library covers its log format and successfully extracts at least one turn with identifiable action and environment-response boundaries; submissions recording only the final patch, or whose logs cannot be segmented into turns, are dropped at this stage. (3)~\textit{Configuration verifiability}: model identity, framework identity and version, and leaderboard resolution label must all be recoverable from directory structure and trajectory headers, so that each trajectory can be attributed to a well-defined $\langle$framework, LLM$\rangle$ configuration. Additional minimum-trajectory filters are applied per configuration at analysis time (\S\ref{sec:per_config_meta}); the resulting per-feature $K$ is reported in \S\ref{sec:evaluation}.

\paragraph{Post-hoc spot check.}
Because parser correctness underpins both inclusion criterion (2) and action classification, we manually audited a 1\% stratified random sample across the 126 configurations, comparing parser output against physical action markers in the raw logs. Any discrepancies were resolved before analysis.

\subsection{Trajectory Type Taxonomy}
\label{par:taxonomy}

\paragraph{Trajectory representation.}
Agent framework design and runtime trace structure are distinct. A single framework can produce structurally different traces depending on the LLM driving it: on mini-swe-agent, weak models terminate in compact sequences while strong models produce extended, exploratory trajectories exceeding 40 turns. Because behavioral analysis operates on traces, we classify \textit{traces}, not designs. We formalize a trajectory as a sequence of turns $\mathcal{T} = \langle t_1, \ldots, t_N \rangle$. Each turn $t_i = \langle \theta_i, a_i, o_i \rangle$ records the model's reasoning trace $\theta_i$ (when exposed by the framework), the executed action $a_i$ (a shell command or tool invocation), and the environment response $o_i$. A trajectory terminates with outcome $Y \in \{\text{resolved}, \text{failed}\}$ as reported by the SWE-bench test suite.

\paragraph{Action classification.}
Each action is classified into one of six semantic categories by intent, regardless of implementation form: a raw \texttt{grep} and a framework-specific code-search API both count as Exploration. A library of 45 format-specific parsers, covering 15 format families, maps 184 distinct action patterns (97 structured tool-call patterns and 87 bash command patterns) into this taxonomy (Table~\ref{tab:action_types}), spanning raw shell commands as well as the custom tool-call formats of SWE-Agent, OpenHands, AutoCodeRover, Honeycomb, Composio, EPAM, and others.

\begin{table*}[!htb]
\caption{Action Classification Schema}
\label{tab:action_types}
\centering
\small
\begin{tabular}{lp{10cm}}
\toprule
\textbf{Category} & \textbf{Representative Actions} \\
\midrule
Exploration & Shell: \texttt{cat}, \texttt{grep}, \texttt{find}, \texttt{ls}; Tool-call: file viewer, code search, fault-localization API \\
Modification & Shell: \texttt{sed}, \texttt{patch}, heredoc; Tool-call: editor commands, structured diff tools \\
Test & \texttt{pytest}, \texttt{unittest} (all frameworks invoke test runners via shell) \\
Navigation & \texttt{cd}, \texttt{pwd} (implicit in tool-call agents) \\
Utility & \texttt{git}, \texttt{pip}, plan selection, submit \\
Unknown & Unclassifiable actions \\
\bottomrule
\end{tabular}
\end{table*}

\noindent The Unknown rate is small in practice: a median of $0.4\%$ per configuration, with 122 of 132 configurations at $\leq 5\%$; only one outlier ($65\%$ unknown) is excluded from action-distribution analyses (\S\ref{sec:evaluation}).

\paragraph{Clustering pipeline.}
To make the 126 configurations legible as a small number of trajectory regimes, we cluster them by how their trajectories actually behave rather than by framework label. Each configuration is summarized by 16 features (Table~\ref{tab:features}) that describe \textit{what} actions its trajectories take, \textit{when} those actions appear, \textit{how} errors propagate, and \textit{how productive} the overall run is. The features are standardized, projected onto three principal components that capture most of the behavioral variance, and grouped by k-means into five trajectory types. The pipeline is deterministic under fixed hyperparameters; exact parameter values, silhouette scores, and full PCA loadings are reported in the replication package. The resulting types serve as the blocking variable for the within-type framework--LLM comparison in RQ1 (\S\ref{sec:rq1}) and as the unit of practitioner guidance in \S\ref{sec:practitioners}.

\paragraph{Feature design rationale.}
Two design choices shape this feature set. First, we aggregate to the configuration level rather than the trajectory level: each $\langle$framework, LLM$\rangle$ configuration runs hundreds of SWE-bench trajectories, so the configuration-level median provides a stable behavioral fingerprint that no single trajectory can provide. Second, each of the four dimensions is captured by several complementary features rather than a single summary statistic. The reason is that single summaries conflate behaviorally distinct regimes: an agent with high error rate may recover quickly or spiral into cascades, but error rate alone cannot separate the two; adding cascade rate, recovery rate, and mean cascade length keeps them separable. The same logic applies across the other dimensions. Action composition uses four mutually exclusive ratios that capture the behaviorally informative categories from Table~\ref{tab:action_types} (Utility and Unknown are excluded from the feature set), and temporal structure pairs a length scalar with transition diversity and four time-anchored features that locate key behavioral transitions in the run. The redundancy within each dimension also makes the clustering robust to noise in any single feature.

\begin{table*}[!htb]
\caption{Sixteen Model-Level Behavioral Features Used for Trajectory Clustering}
\label{tab:features}
\centering
\small
\begin{tabularx}{\linewidth}{ll>{\raggedright\arraybackslash}X}
\toprule
\textbf{Dimension} & \textbf{Feature} & \textbf{Definition (per configuration, aggregated across its trajectories)} \\
\midrule
\multirow{4}{*}{Action composition}
 & Exploration ratio    & Fraction of actions classified as Exploration (Table~\ref{tab:action_types}) \\
 & Modification ratio   & Fraction of actions classified as Modification \\
 & Test ratio           & Fraction of actions classified as Test \\
 & Navigation ratio     & Fraction of actions classified as Navigation \\
\midrule
\multirow{6}{*}{Temporal structure}
 & Trajectory length (log)           & $\log(1+\text{median turns per trajectory})$; log-transformed to compress the heavy tail \\
 & Transition entropy                & Shannon entropy of the action-category transition distribution (diversity of action sequencing) \\
 & Exploration front-loading         & Fraction of Exploration actions occurring in the first 25\% of turns \\
 & First-modification timing         & Normalized turn index of the first Modification action (how early editing begins) \\
 & Phase transition point            & Normalized turn index at which the dominant action category first shifts \\
 & Late-stage entropy                & Transition entropy restricted to the final 25\% of turns (is behavior still exploratory at the end) \\
\midrule
\multirow{5}{*}{Error dynamics}
 & Error rate            & Fraction of turns whose environment response matches one of the 15 error regex categories \\
 & Cascade rate          & Fraction of errors followed by another error within the next 3 turns (error propagation) \\
 & Recovery rate         & Fraction of errors followed by a non-error within the next 3 turns (error resolution) \\
 & Repetition rate       & Fraction of action strings that repeat verbatim within a trajectory (loop/stuck behavior) \\
 & Mean cascade length   & Mean length of consecutive error runs (severity of propagation) \\
\midrule
Efficiency
 & Productive-turn ratio & Fraction of turns that are neither errors nor repeats \\
\bottomrule
\end{tabularx}
\end{table*}

\paragraph{Dimensionality reduction.}
The 16 features are standardized and projected by PCA onto three principal components, which together capture most of the behavioral variance; exact loadings and cumulative variance are reported in the replication package.

\paragraph{Cluster count.}
We use $k=5$, which separates compact retrieval-first traces (which localize before editing) from compact iterative traces (which interleave edits and tests). Silhouette scores marginally favor $k=4$ ($0.40$ versus $0.32$ at $k=5$ and $0.29$ at $k=6$), but $k=4$ merges these two regimes into a single compact cluster and $k=6$ fragments one cluster into a 3-configuration outlier group; we prefer $k=5$ because the retrieval-first versus iterative distinction carries different practitioner implications (\S\ref{sec:practitioners}).
The per-configuration meta-analysis (\S\ref{sec:per_config_meta}) does not take trajectory type as input, so the choice of $k$ affects only the descriptive types in Table~\ref{tab:types} and the within-type comparison in \S\ref{sec:practitioners}: each configuration contributes its own effect size to the meta-analysis independent of which type it falls into, so the headline RQ2/RQ3 quantities are invariant to $k$.

\begin{table}[!htb]
\caption{Trajectory Type Taxonomy (PCA + k-means on 16 behavioral features; T-ent.\ = transition entropy; Front = exploration front-loading ratio)}
\label{tab:types}
\centering
\footnotesize
\begin{tabular}{lrrrr}
\toprule
\textbf{Type} & \textbf{N} & \textbf{Med.\ turns} & \textbf{T-ent.} & \textbf{Front} \\
\midrule
Type~1 \textit{(long-expl.)} & 24 & 44 & 1.25 & 0.18 \\
Type~2 \textit{(iter.-mod.)} & 21 & 37 & 1.16 & 0.42 \\
Type~3 \textit{(broad-div.)} & 48 & 40 & 1.48 & 0.34 \\
Type~4 \textit{(compact-lin.)} &  8 & 10 & 0.29 & 0.32 \\
Type~5 \textit{(compact-front.)} & 25 & 14 & 0.50 & 0.96 \\
\midrule
\textbf{Total} & \textbf{126} & & & \\
\bottomrule
\end{tabular}
\end{table}

\subsubsection{Resulting Types}
Applying the pipeline to the 126 configurations partitions them into five trajectory types (Table~\ref{tab:types}). The types differ along trajectory length, front-loading, and transition entropy (the three columns shown in the table) and span distinct behavioral profiles, which we describe below.

Type~1 (\textit{long-exploratory}) agents have long trajectories with low front-loading: median 44 turns, exploration spread throughout (Front=0.18). Type~2 (\textit{iterative-moderate}) agents have moderate-length trajectories with partial front-loading (37 turns, Front=0.42). Type~3 (\textit{broad-diverse}) agents are the most common profile (48 models, 38\%), with moderate trajectory length (40 turns) and the highest transition entropy (1.48). Type~4 (\textit{compact-linear}) agents are compact and structurally simple: median 10 turns with the lowest entropy (0.29). Type~5 (\textit{compact-frontloaded}) agents are compact and heavily front-loaded: median 14 turns with nearly all exploration concentrated in the first quarter (Front=0.96). The same framework can appear across types: mini-swe-agent produces compact Type~4-like trajectories with weak models and longer, more deliberate ones with strong models, confirming that trajectory type is a property of the trace, not the framework label.

\begin{figure*}[!htb]
\centering
\includegraphics[width=0.85\linewidth]{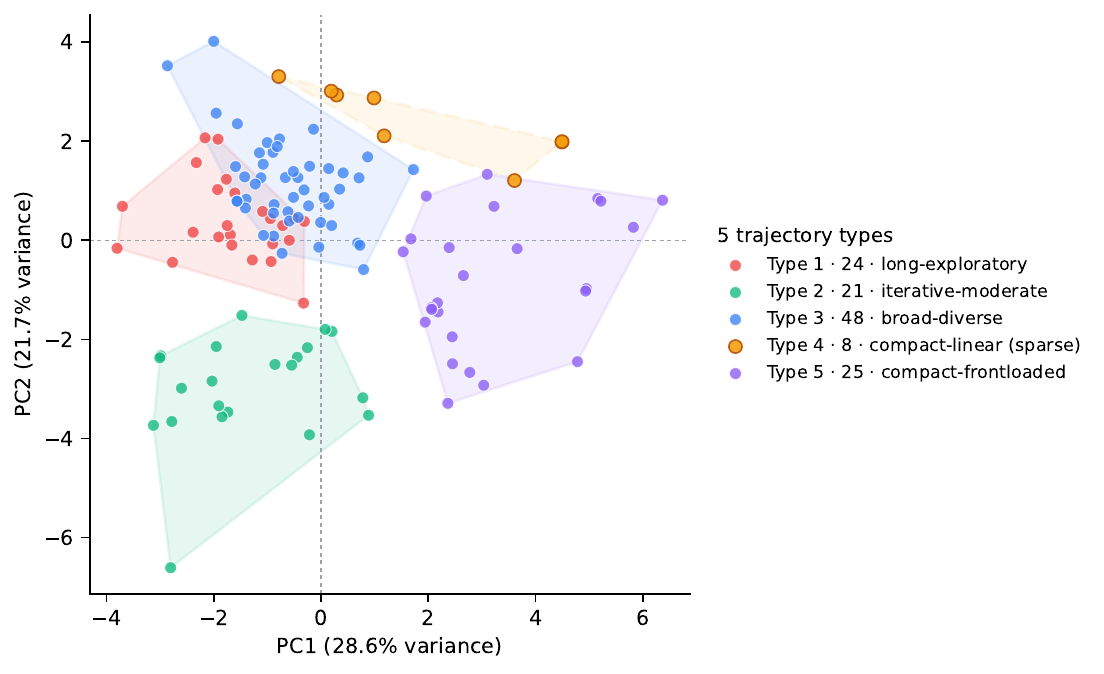}
\caption{The 126 agent configurations projected onto PC1 and PC2 of the 16 behavioral features (cumulative variance $50.3\%$), coloured by trajectory type with convex-hull footprints. Types~1, 2, 3, and 5 form compact, largely separated regions; Type~4 (dashed hull) is sparse, with its 8 members spread along an upper band that touches Type~3 and Type~5 in the projection.}
\label{fig:pca_taxonomy}
\end{figure*}

Figure~\ref{fig:pca_taxonomy} visualizes the 126 configurations in the first two principal components of the 16-feature space, with each type's convex hull drawn behind its points. Types~1, 2, 3, and 5 form compact, largely separated regions consistent with their behavioral descriptions; Type~4 occupies an elongated upper band that is sparse but distinct from the four denser clusters.

\subsection{Control-Flow Graph Features}
\label{par:cfg}

Beyond the sixteen features of Table~\ref{tab:features}, we extract six additional trajectory-level features that describe \textit{how} actions are sequenced rather than only \textit{what} actions occur. They serve two roles in the analyses that follow: a subset of them instantiates candidate behavioral patterns (\S\ref{sec:patterns}), and the full set provides an independent probe that the PCA axes carry behavioral meaning beyond the taxonomy inputs (\S\ref{sec:rq3}).

\subsubsection{Contextual States}
We represent each trajectory as a directed graph $G = \langle V, E \rangle$ built from contextual states. Conditioning each state on more than the action category lets the same action appear as different nodes when its context differs: a retrieval that happens right after an error carries different information than one in the early stage of the trajectory. Every turn $t_i$ is therefore labeled with a state $s_i = \langle \alpha_i, \epsilon_i, \sigma_i \rangle$: the action category $\alpha_i$ (Table~\ref{tab:action_types}); an error context $\epsilon_i \in \{\texttt{clean}, \texttt{post\_error}\}$ marking whether the preceding turn produced an error (\texttt{post\_error}) or not (\texttt{clean}, including the first turn, which has no predecessor); and a trajectory-stage bucket $\sigma_i \in \{\texttt{early}, \texttt{mid}, \texttt{late}\}$ from an equal-thirds split of the turn index. The state alphabet has at most $|S| = 6 \times 2 \times 3 = 36$ labels.

\subsubsection{Graph Construction}
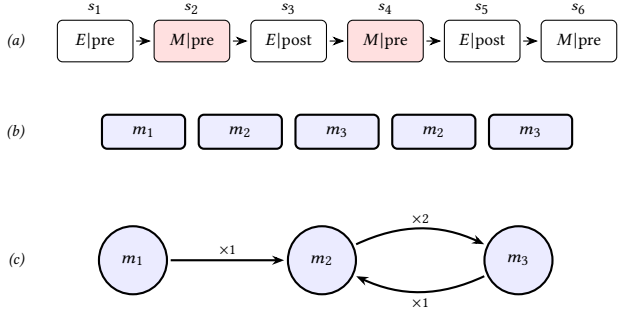
\begin{figure}[!htb]
\centering
\footnotesize
\begin{tikzpicture}[
  >={Stealth[length=1.6mm]},
  st/.style={draw, rounded corners=2pt, minimum width=1.00cm, minimum height=0.55cm, inner sep=1pt, font=\scriptsize},
  err/.style={st, fill=red!12},
  mbox/.style={draw, thick, rounded corners=2pt, fill=blue!7, minimum width=1.1cm, minimum height=0.45cm, inner sep=1pt, font=\scriptsize, align=center},
  cnode/.style={draw, circle, thick, fill=blue!8, minimum size=0.9cm, inner sep=0pt, font=\scriptsize},
  elab/.style={font=\tiny, inner sep=1pt, fill=white},
]
\def\dx{1.28}
\node[st]  (t1) at (0*\dx,0) {$E{\mid}\mathrm{pre}$};
\node[err] (t2) at (1*\dx,0) {$M{\mid}\mathrm{pre}$};
\node[st]  (t3) at (2*\dx,0) {$E{\mid}\mathrm{post}$};
\node[err] (t4) at (3*\dx,0) {$M{\mid}\mathrm{pre}$};
\node[st]  (t5) at (4*\dx,0) {$E{\mid}\mathrm{post}$};
\node[st]  (t6) at (5*\dx,0) {$M{\mid}\mathrm{pre}$};
\foreach \i in {1,...,6}
  \node[above=2pt of t\i, font=\tiny, inner sep=0pt] {$s_{\i}$};
\foreach \i/\j in {t1/t2,t2/t3,t3/t4,t4/t5,t5/t6}
  \draw[->, shorten <=1pt, shorten >=1pt] (\i.east) -- (\j.west);

\foreach \k/\lbl in {0/m_1, 1/m_2, 2/m_3, 3/m_2, 4/m_3} {
  \pgfmathsetmacro{\xp}{\k*\dx + 0.5*\dx}
  \node[mbox] at (\xp,-1.2) {$\lbl$};
}

\node[font=\scriptsize\itshape, anchor=east] at (-0.85,0)    {(a)};
\node[font=\scriptsize\itshape, anchor=east] at (-0.85,-1.2) {(b)};

\def\cy{-2.9}
\node[cnode] (M1) at (0.5,\cy) {$m_1$};
\node[cnode] (M2) at (3.0,\cy) {$m_2$};
\node[cnode] (M3) at (5.6,\cy) {$m_3$};
\draw[->, thick, shorten <=1pt, shorten >=1pt] (M1) -- node[elab, above=1pt] {$\times 1$} (M2);
\draw[->, thick, bend left=25, shorten <=1pt, shorten >=1pt] (M2) to node[elab, above=1pt] {$\times 2$} (M3);
\draw[->, thick, bend left=25, shorten <=1pt, shorten >=1pt] (M3) to node[elab, below=1pt] {$\times 1$} (M2);
\node[font=\scriptsize\itshape, anchor=east] at (-0.85,\cy) {(c)};
\end{tikzpicture}
\caption{CFG construction on a 6-turn example (stage bucket $\sigma$ omitted for readability). \textbf{(a)}~Each turn $t_i$ carries state $s_i = \langle \alpha_i, \epsilon_i, \sigma_i \rangle$, with $\alpha_i$ the action category ($E$=exploration, $M$=modification) and $\epsilon_i \in \{\mathrm{pre}, \mathrm{post}\}$ the error context. Shaded turns $t_2, t_4$ emitted environment errors, so their successors $s_3, s_5$ carry $\epsilon = \mathrm{post}$. \textbf{(b)}~Each motif $m_i = \langle s_i, s_{i+1} \rangle$ is an adjacent state pair; five motif instances collapse to three distinct motifs, with $m_2$ and $m_3$ each recurring once. \textbf{(c)}~The distinct motifs form $V$; four edge instances group into three directed edges with the shown multiplicities. Derived features for this trajectory: revisit rate $= 2/5 = 0.40$, backtrack rate $= 2/3 \approx 0.67$ (positions $t=4, 5$ satisfy $m_t = m_{t-2}$), self-loop rate $= 0$, and post-error motif ratio $= 4/5 = 0.80$.}
\label{fig:cfg_example}
\end{figure}

Nodes and edges are defined at the \textit{motif} level. A motif $m_i = \langle s_i, s_{i+1} \rangle$ is an ordered pair of consecutive state labels, the minimal unit that records adjacency in the state sequence. The \textit{node set} $V$ contains the distinct motifs observed in the trajectory; the 2-step choice bounds $|V|$ by $|S|^2$ and keeps $V$ small enough ($\sim\!$ a few dozen motifs per trajectory in practice) to support stable entropy and revisit statistics. The \textit{edge set} $E$ contains a directed edge $m_i \to m_{i+1}$ for every pair of consecutive motif instances, with edge multiplicity counting how often each motif pair is observed. A trajectory of $N$ turns produces $N-1$ motif instances and $N-2$ edge instances; because $m_i$ and $m_{i+1}$ share the state $s_{i+1}$, an edge in $G$ encodes a 3-turn sliding window over the state sequence. Figure~\ref{fig:cfg_example} illustrates the construction on a small worked example.

\subsubsection{Derived Features}
The six features computed over $G$ target three behavioral phenomena: (i) action-sequence diversity, (ii) persistent repetition or oscillation, and (iii) operation under recent-error context.

For diversity, \textit{motif entropy} and \textit{transition entropy} are the Shannon entropies of the motif-instance distribution over $V$ and the edge-instance distribution over $E$, respectively; higher values indicate more diverse 2-turn and 3-turn patterns.
For repetition and loops, \textit{self-loop rate} is the fraction of edges whose source and target motifs are identical (three consecutive identical states); \textit{revisit rate} is the fraction of motif instances whose motif had already appeared earlier in the trajectory; \textit{backtrack rate} is the fraction of motif positions $t \geq 3$ satisfying $m_t = m_{t-2}$, capturing short-range oscillations.
For error context, \textit{post-error motif ratio} is the fraction of motif instances in which at least one state carries $\epsilon = \texttt{post\_error}$.

\subsection{Candidate Behavioral Patterns}
\label{sec:patterns}

We define seven binary behavioral patterns (P1--P7) for testing whether specific behaviors predict task resolution. Each pattern is operationalized as a binary indicator (present/absent per trajectory), with the threshold derived from the bash-only subset and validated on the held-out verified subset. The seven patterns fall along three dimensions of agent behavior: action sequencing, error recovery, and trajectory efficiency; Table~\ref{tab:candidate_patterns} lists all seven.

The seven patterns are constructed by reviewing prior single-framework SE-agent studies~\cite{shepherd2025, chen2026process, bouzenia2025trajectory, majgaonkar2026trajectory} and classical debugging practice~\cite{zeller2009}, extracting behaviorally distinct rules operationalizable as binary indicators. Each pattern below cites either its direct source (P1, P3, P5, P7) or the sources whose findings motivate its synthesis (P2, P4, P6). Thresholds are either fixed by design (P2, P4) or set to the median over the bash-only subset (P3, P5, P6), with the latter re-tested in \S\ref{sec:threats}.

\paragraph{Action sequencing (P1, P2, P7).}
These test the order and balance of agent actions during problem-solving. P1 (exploration before the first modification) reflects the classical debugging principle that developers should understand code before changing it~\cite{zeller2009}. P7 (at least one test after a modification) aligns with test-driven development (TDD) practice and captures one manifestation of SHEPHERD's \emph{False Termination}, prematurely assuming task completion~\cite{shepherd2025}. P2 (exploration ratio $\in [.30, .50]$) operationalizes the action-balance concern raised by Bouzenia and Pradel~\cite{bouzenia2025trajectory}: insufficient exploration risks premature termination, excessive exploration risks repetitive cycles. Sensitivity analysis (\S\ref{sec:threats}) confirms classification stability under plausible variations.

\paragraph{Error recovery (P3, P4).}
These capture how the agent handles errors when they arise. P3 (longest error cascade below the bash-only median) is motivated by Chen et al.'s finding that higher error frequency correlates with lower resolution rates~\cite{chen2026process}; we use cascade length rather than raw error count to isolate recovery quality. P4 (first cascade resolved within $\leq 2$ turns) operationalizes recovery from the \textit{repetitive cycle} anti-pattern identified by Bouzenia and Pradel~\cite{bouzenia2025trajectory}: agents that exit their first cascade quickly avoid the high overall error rates that Chen et al. associate with low resolution~\cite{chen2026process}. The 2-turn cutoff is stable under sensitivity analysis.

\paragraph{Trajectory efficiency (P5, P6).}
These measure the overall shape of a trajectory independent of its content. P5 (length below median) follows Majgaonkar et al.'s finding that failed trajectories are consistently longer than resolved ones~\cite{majgaonkar2026trajectory}, operationalized as a binary indicator at the bash-only median. P6 (late-stage entropy below median) tests whether late-trajectory action diversity signals resolution: low late entropy can indicate either convergent problem-solving or the \textit{repetitive cycle} anti-pattern identified by Bouzenia and Pradel~\cite{bouzenia2025trajectory}, where the agent gets stuck repeating the same actions. Both P5 and P6 are thresholded at the bash-only median, and sensitivity analysis (\S\ref{sec:threats}) shows their classification is stable across alternative thresholds.

\begin{table}[!htb]
\caption{Candidate Behavioral Patterns (P1--P7)}
\label{tab:candidate_patterns}
\centering
\footnotesize
\setlength{\tabcolsep}{3pt}
\begin{tabular}{clp{3.2cm}}
\toprule
& \textbf{Name} & \textbf{Criterion} \\
\midrule
P1 & Navigate-before-modify & Exploration precedes first Modification \\
P2 & Moderate exploration  & Expl.\ ratio $\in [.30,.50]$ \\
P3 & Short cascades        & Max cascade $<$ median \\
P4 & Fast cascade recovery & 1st cascade $\leq 2$ turns \\
P5 & Short trajectory      & Length $<$ median \\
P6 & Low late entropy      & Late-stage entropy $<$ median \\
P7 & Test-after-modify     & Test follows $\geq\!1$ Modification \\
\bottomrule
\end{tabular}
\end{table}

\subsection{Statistical Methods}
\label{par:stats}
\paragraph{Two-level structure.}
The analyses operate at two scales. \textit{Within} each configuration, a per-configuration effect size quantifies the strength of association between a behavioral feature and task resolution among that configuration's trajectories. \textit{Across} configurations, random-effects meta-analysis~\cite{borenstein2009meta} aggregates the per-configuration effects and tests whether they transfer to the ecosystem or disagree (\S\ref{sec:per_config_meta}).

\paragraph{Effect-size estimators.}
Behavioral metrics are non-normal: bounded ratios, heavy-tailed counts, and lengths spanning three orders of magnitude. We therefore use rank-based effect sizes throughout, which also makes the estimates unit-agnostic across loop-based and pipeline frameworks. For RQ2 binary patterns (P1--P7) we use signed Cram\'er's $V$~\cite{cramer1946methods}, the standard chi-square-based association strength on the $2 \times 2$ table of (pattern present/absent) $\times$ (resolved/failed). For RQ3 continuous features we use the rank-biserial correlation~\cite{kerby2014simple}
\[
r \;=\; 1 - \frac{2U}{n_1 n_2},
\]
where $U$ is the Mann--Whitney rank-sum statistic~\cite{mann1947test} and $n_1, n_2$ are the resolved and unresolved trajectory counts in the configuration. Both $V$ and $r$ are bounded between $-1$ and $+1$, with sign aligned so that positive values mean the behavior co-occurs with resolution: pattern presence for binary patterns, or lower feature values for continuous features. For the RQ1 cross-configuration Kruskal--Wallis comparisons~\cite{kruskal1952ranks} we report $\eta^2$, the fraction of feature rank variance explained by the grouping factor; we adopt Cohen's threshold $\eta^2 \geq 0.14$ for a large effect~\cite{cohen1988statistical}.

 We report effect sizes and direction splits throughout, rather than null-hypothesis p-values. With $n = 64{,}380$ trajectories any non-zero association crosses any conventional significance threshold by virtue of sample size alone~\cite{sullivan2012using, wasserstein2016asa}, and our central claims concern cross-configuration heterogeneity rather than non-zeroness of a pooled effect, so the $I^2$ and direction-splits framework is the appropriate inference target~\cite{borenstein2009meta}.

\subsubsection{Per-configuration meta-analysis}
\label{sec:per_config_meta}
\paragraph{Motivation.}
Pooling all 64,380 trajectories into a single regression would conflate within-configuration signal with between-configuration heterogeneity, which is exactly the confusion this paper aims to resolve. We instead adopt the classical random-effects meta-analysis formulation~\cite{dersimonian1986meta, borenstein2009meta} at the configuration level: each of the 126 configurations plays the role of one ``study'' in the meta-analytic vocabulary, contributing its own effect size, and we test whether those per-configuration effects agree across configurations.

\paragraph{Per-configuration effect sizes.}
Let $V_{im}$ denote the effect size of feature $i$ on resolution within configuration $m$, computed with the rank-based estimators above ($V$ for binary patterns, $r$ for continuous features); the pooled mean is correspondingly written $\bar V$ in binary-pattern tables (Table~\ref{tab:patterns}) and $\bar r$ in continuous-feature tables (Table~\ref{tab:behavior_outcome}), reflecting which underlying effect size is being averaged. We compute $V_{im}$ for every configuration with at least 20 trajectories, of which at least 5 are resolved and at least 5 are unresolved. A configuration is dropped from the meta-analysis of feature $i$ when its within-configuration effect is undefined; for example, a binary pattern that is saturated within a configuration makes Cram\'er's $V$ degenerate. We write $K_i$ for the number of configurations that contribute a valid $V_{im}$ to feature $i$. This count is feature-specific: $K_i = 119$ for every continuous RQ3 feature (only the 7 of 126 configurations with extreme resolution rates are dropped), while for binary patterns $K_i$ ranges from $40$ (P1, most often saturated) to $110$ (P3). Exact per-feature values appear in Tables~\ref{tab:patterns} and~\ref{tab:behavior_outcome}.

\paragraph{Aggregating with $I^2$.}
We pool the $K_i$ valid per-configuration estimates under the random-effects model of DerSimonian and Laird~\cite{dersimonian1986meta}. Each configuration is assigned an inverse-variance weight $w_{im} = \hat{\sigma}_{im}^{-2}$, where $\hat{\sigma}^2_{im}$ is the estimated sampling variance of $V_{im}$, so configurations with more trajectories carry more weight. The meta-analytic point estimate is the weighted pooled mean
\[
\bar{V}_i \;=\; \dfrac{\displaystyle\sum_{m} w_{im}\, V_{im}}{\displaystyle\sum_{m} w_{im}},
\]
and Cochran's $Q$ statistic~\cite{cochran1954combination}
\[
Q_i \;=\; \sum_{m} w_{im}\,(V_{im} - \bar{V}_i)^2
\]
measures how much the per-configuration effects deviate from this pooled mean. The $I^2$ statistic of Higgins and Thompson~\cite{higgins2002quantifying, higgins2003measuring} rescales $Q_i$ to a fraction:
\[
I^2_i \;=\; \max\!\bigl(0,\; 1 - (K_i - 1)/Q_i\bigr) \times 100\%,
\]
where $K_i - 1$ is the value of $Q$ expected under the null of universal agreement, so $1 - (K_i - 1)/Q_i = 0$ exactly when the observed dispersion matches what sampling noise alone would produce. The $\max(\cdot)$ clips the rare negative values that arise when sampling noise over-explains the observed variation. Intuitively, $I^2$ is the share of cross-configuration variation that sampling noise alone cannot explain: $I^2 = 0\%$ means configurations agree up to noise, and $I^2 = 100\%$ means their effects genuinely differ. Following Higgins et al.~\cite{higgins2003measuring}, we classify configurations as universal ($I^2 < 25\%$), moderate ($25\%$--$75\%$), or configuration-specific ($I^2 \geq 75\%$).

\paragraph{Direction splits.}
$I^2$ captures the \textit{magnitude} of cross-configuration disagreement but not its \textit{direction}, and standard meta-analytic practice complements it with sign-based diagnostics~\cite{borenstein2009meta, higgins2003measuring}. An $I^2$ of $80\%$, for example, is compatible with two very different regimes: configurations all pushing the same way with consistently large effects (large but agreeing), or splitting between strictly positive and strictly negative effects (genuinely contradictory). We report this sign distribution explicitly as $\langle n_+, n_- \rangle$ alongside $I^2$ for every feature (written $n_+/n_-$ in tables): the number of configurations with a strictly positive versus strictly negative per-configuration effect, with near-zero effects excluded from both counts. A roughly balanced split (e.g., $47/48$) is direct evidence that the same signal plays opposite roles across configurations, and unlike $I^2$ this conclusion does not depend on estimation precision: the count is a deterministic function of the sign pattern.

\paragraph{Moderators via meta-regression.}
To identify which layer of the design stack drives the observed heterogeneity, we extend the random-effects model with \emph{moderators}~\cite{borenstein2009meta, raudenbush2009meta}: configuration-level variables (specifically, framework identity and LLM family) that may explain why per-configuration effect sizes differ. In meta-analytic terms, a moderator is a study-level covariate intended to explain part of the between-configuration true-effect variance $\tau^2$ (i.e., the dispersion of $V_{im}$ remaining after the sampling-error component is subtracted). For each feature we fit two \emph{separate} meta-regressions: one with framework identity (43 levels) as the sole moderator, one with LLM family (6 levels) as the sole moderator. We use separate fits rather than a single joint regression because $K_i \leq 119$ valid configurations cannot stably estimate $43 + 6 = 49$ moderator indicators in one fit; separate fits keep each regression well-posed. For each fit we record the meta-analytic pseudo-$R^2$ of Raudenbush~\cite{raudenbush2009meta}:
\[
R^2 \;=\; 1 - \frac{\tau^2_{\text{residual}}}{\tau^2_{\text{null}}} \;\in\; [0,\,1],
\]
where $\tau^2_{\text{null}}$ is $\tau^2$ in the null random-effects model with no moderator, and $\tau^2_{\text{residual}}$ is the $\tau^2$ that remains after the moderator is added. $R^2$ is therefore the share of cross-configuration true-effect variance that the moderator alone explains. We denote the two fits $R^2_{\text{FW}}$ and $R^2_{\text{LLM}}$. Because they are fit independently, $R^2_{\text{FW}} + R^2_{\text{LLM}}$ can exceed $1$ when framework and LLM share variance; we therefore report the \textit{ratio} $R^2_{\text{FW}}\,/\,R^2_{\text{LLM}}$ as the framework-versus-LLM dominance indicator (ratios well above $1$ mean framework dominates, ratios near $1$ mean shared explanatory power, ratios well below $1$ mean LLM dominates).

\paragraph{Robustness diagnostics.}
The point estimate of $R^2_{\text{FW}}$ from a 43-level dummy regression has two weaknesses that single-number reporting obscures. First, linear regression with high-cardinality categorical moderators overfits sampling noise: shuffling framework labels among configurations and recomputing $R^2$ on the shuffled data yields a chance baseline with mean $0.33$ and 95th percentile $0.45$ across the 13 continuous features, so an observed $R^2_{\text{FW}}$ below $0.45$ is not distinguishable from a random label assignment. Second, the point estimate alone reveals neither its precision nor whether a single framework dominates the fit. We therefore accompany each headline $R^2$ with three diagnostics: (i) a 95\% bootstrap percentile interval over 2{,}000 resamples of configurations, for precision; (ii) a one-sided permutation null over 2{,}000 random framework-label assignments, for the chance baseline, and we count the framework moderator as significant for a feature when its observed $R^2_{\text{FW}}$ exceeds the null 95th percentile (permutation $p < 0.05$); and (iii) exhaustive leave-one-framework-out resampling, for sensitivity to any single framework. \S\ref{sec:robustness} reports the per-feature outcomes; headline claims about framework as moderator rely only on features that clear the chance baseline.

\subsubsection{Framework Effect Isolation}
Separating framework effects from LLM effects requires varying one factor while holding the other fixed. This is a natural experiment in the epidemiological sense~\cite{craig2017natural}. The SWE-bench leaderboard provides such an opportunity: several LLMs appear across many frameworks, and we identify three that each appear in $\geq$5 distinct frameworks (Table~\ref{tab:tracers}). Because each tracer is pinned to a single exact model version, behavioral differences within a tracer reflect framework design rather than model variation.

\begin{table}[!htb]
\caption{LLM Tracers for Framework Effect Isolation}
\label{tab:tracers}
\centering
\small
\begin{tabular}{lccr}
\toprule
\textbf{Tracer} & \textbf{Version} & \textbf{Frameworks} & \textbf{Trajectories} \\
\midrule
Claude 4 Sonnet & 20250514 & 8 & 3,985 \\
Claude 3.5 Sonnet & 20241022 & 8 & 5,414 \\
GPT-4o & 2024 & 6 & 3,459 \\
\bottomrule
\end{tabular}
\end{table}

Each tracer is an exact model version (e.g., \texttt{claude-\allowbreak sonnet-\allowbreak 4-\allowbreak 20250514}); different submissions of the same version are merged into one tracer. The two Claude tracers (3.5 and 4) provide a cross-generation consistency check: framework effects that persist across both tracers reflect structural framework properties rather than model-specific interactions. RQ1 uses this check to distinguish robust framework effects from generation-specific artifacts. The framework-effect decomposition is a bundle treatment: action interface, tool design, system prompt, and configuration choices are co-varied by the design and not individually isolated (\S\ref{sec:threats}).

%% file: sections/evaluation.tex
\section{Evaluation Setup}
\label{sec:evaluation}

\paragraph{Data Sources and Collection.}
Trajectories come from two sources: (1)~the SWE-bench Experiments repository\footnote{\url{https://github.com/SWE-bench/experiments}}, which hosts leaderboard submissions organized by subset and model, and (2)~the Docent platform\footnote{\url{https://docent.transluce.org/}}, which provides full conversation transcripts with thoughts, actions, and observations. All trajectories target SWE-bench Verified~\cite{jimenez2024swebench} under the 100\% oracle setting (full test visibility). SWE-bench Verified hosts cross-framework leaderboard submissions at the scale required for our analysis. Raw logs are parsed by 45 format-specific parsers (15 format families) into structured (thought, action, observation) triples.

\paragraph{Two-dataset design.}
The \textit{bash-only} subset (16,522 trajectories, 33 LLMs on mini-swe-agent) holds the framework constant, isolating LLM-level variation; it serves as the controlled baseline for RQ1 and the pattern-definition reference for RQ2. The \textit{verified} subset (47,858 trajectories, 42 frameworks, 93 configurations) supplies the framework diversity required for the per-configuration meta-analysis across 126 configurations. Together the two subsets cover 64,380 trajectories from 126 configurations across 43 frameworks.

\paragraph{Dataset statistics.}
Table~\ref{tab:stats_by_arch} reports trajectory counts and resolution rates by type. Type~1 and Type~4 share the lowest resolution rate (42.1\%), Type~2 the highest (57.6\%). The five types are defined by data-driven clustering of trace properties; the same framework may appear in multiple types depending on which LLM configuration is used.

\begin{table}[!htb]
\centering
\caption{Trajectories and Resolution Rates by Trajectory Type}
\label{tab:stats_by_arch}
\small
\begin{tabular}{lrrr}
\toprule
\textbf{Type} & \textbf{Configs} & \textbf{Traj.} & \textbf{Res.\%} \\
\midrule
Type~1 & 24 & 13,688 & 42.1 \\
Type~2 & 21 & 10,500 & 57.6 \\
Type~3 & 48 & 23,737 & 49.8 \\
Type~4 &  8 &  3,981 & 42.1 \\
Type~5 & 25 & 12,474 & 46.3 \\
\midrule
\textbf{Total} & \textbf{126} & \textbf{64,380} & 48.3 \\
\bottomrule
\end{tabular}
\end{table}

\paragraph{Inclusion criteria.}
The single configuration with $>$50\% unknown actions is excluded from action-distribution analyses but retained for resolution-rate comparisons. For the per-configuration meta-analysis (RQ2/RQ3), a configuration contributes an effect size only if it has sufficient resolved and unresolved trajectories for the feature in question (typically $K=40$--$119$ of 126 configurations per feature); configurations with degenerate splits are dropped per-feature. These filters yield 126 configurations overall.

\paragraph{Data quality.}
Classifier accuracy was validated by manual annotation of 500 randomly sampled turns across 5 frameworks (Cohen's $\kappa > 0.85$). Format detection is unambiguous: on a random sample of 500 log files, each file matched the format signature of exactly one of the 45 parsers, so no trajectory is attributable to more than one format family. For models available from both data sources, parsed turn counts and resolution labels were cross-verified, and discrepancies were resolved by preferring the more complete source.

%% file: sections/results.tex
\section{Results}
\label{sec:results}

\subsection{RQ1: Framework vs.\ LLM}
\label{sec:rq1}

\textbf{RQ1}: \textit{What drives behavioral differences across agent configurations: framework design or LLM capability?}

\paragraph{Setup.}
RQ1 uses two complementary analyses on the multi-tracer slice (\S\ref{sec:design}, three tracers each appearing in 6--8 frameworks) and the bash-only baseline. The \textit{multi-tracer analysis} estimates framework effects within each tracer via Kruskal--Wallis $\eta^2$ on the behavioral features defined in Table~\ref{tab:features}, with paired Wilcoxon signed-rank tests on matched instances (same task, same LLM, different framework) as a task-difficulty control. The \textit{within-type controlled comparison} holds trajectory type fixed so that framework and LLM contributions are measured on a like-for-like trace-structure baseline: \textit{framework variance} (FW-Var) is the Kruskal--Wallis $\eta^2$ across frameworks within the same (type, tracer) cell, and \textit{LLM variance} (LLM-Var) is the Kruskal--Wallis $\eta^2$ across LLM families within mini-swe-agent at fixed framework.

The multi-tracer analysis shows large framework effects within each tracer. All 27 tracer $\times$ action-feature tests are significant ($p < 0.001$, $\eta^2 = 0.17$--$0.90$; Table~\ref{tab:traj_tracer}), with median $\eta^2$ of 0.78 (Claude~4 Sonnet), 0.33 (Claude~3.5 Sonnet), and 0.56 (GPT-4o). Action ratios are computed from raw actions, not framework labels. Paired Wilcoxon tests (same task, same LLM, different framework) show 84.5\%--95.5\% of comparisons significant at $p < 0.001$, confirming the effect persists under task-difficulty control across 71 framework pairs that each share at least 300 SWE-bench Verified tasks (Jaccard $\approx 1.0$).

\begin{table}[!htb]
	\centering
	\caption{Framework Effect on Action Features (Tracer Analysis)}
	\label{tab:traj_tracer}
	\small
	\begin{tabular}{lrrr}
		\toprule
		\textbf{Tracer} & \textbf{$n$} & \textbf{Med.\ $\eta^2$} & \textbf{Paired sig.\ ($p{<}0.001$)} \\
		\midrule
		Claude 4 Sonnet   & 3,985 & 0.78 & 94.3\% \\
		Claude 3.5 Sonnet & 5,414 & 0.33 & 84.5\% \\
		GPT-4o            & 3,459 & 0.56 & 95.5\% \\
		\bottomrule
	\end{tabular}
\end{table}

A controlled within-type comparison (same trajectory type, same tracer LLM, FW-Var vs.\ LLM-Var) across 63 (type, tracer, feature) triples shows FW-Var $>$ LLM-Var in 44 of 63 cases (69.8\%; Table~\ref{tab:dir_ab}). The pattern is type-conditioned. In Type~3 (\textit{broad-diverse}) and Type~5 (\textit{compact-frontloaded}), framework variance dominates strongly: FW win rates 94\% and 83\%, with median FW $\eta^2 = 0.48$ and $0.36$ versus LLM $\eta^2 = 0.11$ and $0.02$. Type~2 is also framework-leaning (FW wins 6/9, 67\%; median FW $\eta^2 = 0.22$ vs.\ LLM $\eta^2 = 0.02$). Type~1 (\textit{long-exploratory}) is the consistent exception: LLM-Var exceeds FW-Var in 12 of 18 comparisons (67\%), with median LLM $\eta^2 = 0.28$ vs.\ FW $\eta^2 = 0.08$. Type~4 is not covered (insufficient same-type tracer pairs across its 4 frameworks).

\begin{table*}[!htb]
\caption{Within-Type Controlled Variance Decomposition: FW-Var vs.\ LLM-Var (same trajectory type, same tracer LLM; 63 (type, tracer, feature) triples across 4 covered types)}
\label{tab:dir_ab}
\centering
\small
\begin{tabular}{lcccc}
\toprule
\textbf{Type} & \textbf{FW wins} & \textbf{Med.\ FW $\eta^2$} & \textbf{Med.\ LLM $\eta^2$} & \textbf{Dominant} \\
\midrule
Type~1 &  6/18 (33\%) & 0.08 & 0.28 & \textbf{LLM} \\
Type~2 &  6/9  (67\%) & 0.22 & 0.02 & Framework \\
Type~3 & 17/18 (94\%) & 0.48 & 0.11 & Framework \\
Type~5 & 15/18 (83\%) & 0.36 & 0.02 & Framework \\
\midrule
\textbf{Overall} & \textbf{44/63 (70\%)} & & & Framework \\
\bottomrule
\end{tabular}
\end{table*}

\begin{rqbox}[Answer to RQ1]
Both framework and LLM substantially shape agent behavior; framework identity is the larger driver in most regimes, but the balance is type-conditioned. FW-Var $>$ LLM-Var in 44 of 63 within-type comparisons (69.8\%); multi-tracer Kruskal--Wallis $\eta^2 = 0.17$--$0.90$ for framework effects on action features. Framework dominates in Type~2, Type~3, and Type~5 (FW win rates 67\%--94\%); Type~1 (long-exploratory) is the consistent exception with LLM dominance (12 of 18 comparisons); Type~4 is not covered (insufficient same-type tracer pairs).
\end{rqbox}

\subsection{RQ2: Pattern Universality}
\label{sec:rq2}

\textbf{RQ2}: \textit{Which behavioral rules transfer across agent configurations, and which are configuration-specific?}

\paragraph{Setup.} We test 7 binary behavioral patterns (P1--P7; Table~\ref{tab:candidate_patterns}) at the per-configuration level (\S\ref{sec:per_config_meta}): for each configuration with $\geq$20 trajectories, $\geq$5 resolved, and $\geq$5 unresolved, we compute signed Cram\'er's $V$ on its trajectories, then aggregate to a per-pattern $I^2$ across the $K$ valid configurations. Framework identity (43 levels) and LLM family (6 levels) enter as moderators in random-effects meta-regression, yielding $R^2_{\text{FW}}$ and $R^2_{\text{LLM}}$. Classification follows Higgins et al.: universal ($I^2 < 25\%$), moderate ($25\%$--$75\%$), configuration-specific ($\geq 75\%$).

Table~\ref{tab:patterns} reports the per-pattern diagnostics: $K$ valid configurations, weighted effect $\bar V$, direction split $n_+/n_-$, heterogeneity $I^2$, and moderator $R^2$ for framework identity and LLM family.

\begin{table*}[!htb]
	\begin{threeparttable}
		\caption{RQ2 Per-Configuration Meta-Analysis: Seven Binary Patterns Across 126 Agent Configurations}
		\label{tab:patterns}
		\centering
		\small
		\begin{tabular}{llrcrrrrl}
			\toprule
			\textbf{ID} & \textbf{Pattern} & \textbf{$K$} & \textbf{$\bar V$} & \textbf{$n_+/n_-$} & \textbf{$I^2$} & \textbf{$R^2_{\text{FW}}$} & \textbf{$R^2_{\text{LLM}}$} & \textbf{Class.} \\
			\midrule
			P5 & Shorter trajectory      & 103 & $+0.153$ & 93/8  & \textbf{86.8\%} & 31.6\% & 11.4\% & Config-spec. \\
			P7 & Test-after-modify       &  86 & $+0.017$ & 50/33 & \textbf{75.0\%} & 38.1\% & 37.8\% & Config-spec. \\
			P6 & Lower late entropy      & 101 & $-0.008$ & 45/46 & 69.7\%          & 47.6\%$^\dagger$ & 33.8\% & Moderate \\
			P2 & Moderate exploration    & 106 & $+0.022$ & 60/34 & 68.9\%          & 44.0\% & 21.9\% & Moderate \\
			P4 & Fast cascade recovery   & 111 & $+0.001$ & 44/54 & 59.4\%          & 25.2\% & 26.8\% & Moderate \\
			P1 & Navigate-before-modify  &  40 & $+0.026$ & 26/12 & 57.3\%          & 58.9\% & 46.1\% & Moderate \\
			P3 & Shorter cascades        & 108 & $+0.055$ & 86/13 & 50.8\%          & 37.5\% & 21.3\% & Moderate \\
			\bottomrule
		\end{tabular}
		\begin{tablenotes}
			\footnotesize
			\item $K$: valid configurations after minimum-trajectory filters. $\bar V$: inverse-variance-weighted mean of signed Cram\'er's $V$ (positive = pattern presence associated with resolution). $n_+/n_-$: number of configurations with positive vs.\ negative per-configuration effect (residual $K - n_+ - n_-$ are zero-effect). $R^2_{\text{FW}}$ ($R^2_{\text{LLM}}$): fraction of between-configuration variance in $V$ explained by framework identity (LLM family) in a random-effects meta-regression. Classification: universal ($I^2 < 25\%$), moderate ($25$--$75\%$), configuration-specific ($\geq 75\%$). $\dagger$ marks $R^2_{\text{FW}}$ values that exceed the framework permutation null at $\alpha = 0.05$ (\S\ref{sec:robustness}).
		\end{tablenotes}
	\end{threeparttable}
\end{table*}

\paragraph{Two patterns are configuration-specific, with different signatures.}
P5 (shorter trajectory, $I^2 = 86.8\%$, $K = 103$) has the largest weighted effect ($\bar V = +0.153$) and is predominantly positive (93 of 103 configurations), but individual values span $V = -0.36$ to $V = +0.40$: heterogeneity is in magnitude, not direction. A framework-aware calibration of \textit{how much shorter} is therefore required even though the direction holds. P7 (test-after-modify, $I^2 = 75.0\%$, $K = 86$) sits on the moderate/configuration-specific boundary and is qualitatively different: the weighted effect is small ($\bar V = +0.017$) and the direction itself disagrees across configurations (50 positive, 33 negative).

\paragraph{Five patterns show moderate heterogeneity.}
The other five fall in the moderate band ($I^2 = 50.8\%$--$69.7\%$). Two reverse direction across configurations: P4 (fast cascade recovery, 44/54) and P6 (low late entropy, 45/46) are nearly balanced, with the rule pointing opposite ways for different configurations.

\paragraph{Only P6 clears the chance baseline.}
Numerically, framework $R^2$ exceeds LLM $R^2$ for five of seven patterns ($R^2_{\text{FW}}/R^2_{\text{LLM}}$ from $1.3\times$ to $2.8\times$). Under the permutation diagnostic in \S\ref{sec:robustness}, however, only P6 (low late entropy) clears the chance baseline at $\alpha = 0.05$ (Table~\ref{tab:patterns}, $\dagger$); the binary patterns saturate to highly skewed distributions ($n_+/n_-$ heavily one-sided in five of seven), which limits meta-regression power on this discretized signal.

\begin{rqbox}[Answer to RQ2]
\textbf{No binary behavioral pattern is universal} across the 126 configurations. Two of seven (P5 shorter-trajectory, P7 test-after-modify) are configuration-specific ($I^2 \geq 75\%$); the remaining five are moderate ($I^2 = 51$--$70\%$). Three patterns (P4, P6, P7) reverse direction across configurations, so the same rule points opposite ways depending on the target configuration. Numerically framework $R^2$ exceeds LLM $R^2$ for five of seven patterns, but only P6 clears the permutation chance baseline (\S\ref{sec:robustness}).
\end{rqbox}

\subsection{RQ3: Behavioral Predictors}
\label{sec:rq3}

\textbf{RQ3}: \textit{Does the predictive value of behavioral features depend on agent configuration, and which layer of the stack drives that dependence?}

\paragraph{Setup.}
We test 13 continuous features: 7 trajectory-level (3 action ratios: exploration, modification, test; 3 error metrics: error rate, cascade rate, repetition rate; and mean turns) and 6 CFG-structural (\S\ref{sec:design}). For each (feature, configuration) pair, Mann--Whitney $U$ compares resolved versus unresolved trajectories; rank-biserial $r = 1 - 2U/(n_1 n_2)$ is the per-configuration effect size, with positive $r$ meaning lower feature value associates with resolution. $I^2$ is computed across the 119 configurations meeting the minimum-trajectory filter. Framework identity and LLM family enter as moderators in random-effects meta-regression, yielding $R^2_{\text{FW}}$ and $R^2_{\text{LLM}}$.

\begin{table*}[!htb]
\begin{threeparttable}
\caption{RQ3 Per-Configuration Meta-Analysis: Thirteen Continuous Features Across 119 Agent Configurations}
\label{tab:behavior_outcome}
\centering
\small
\begin{tabular}{llrcrrrrl}
\toprule
& \textbf{Feature} & \textbf{$K$} & \textbf{$\bar r$} & \textbf{$n_+/n_-$} & \textbf{$I^2$} & \textbf{$R^2_{\text{FW}}$} & \textbf{$R^2_{\text{LLM}}$} & \textbf{Class.} \\
\midrule
\multicolumn{9}{l}{\textit{Action features}} \\
 & mean turns         & 119 & $+0.320$ & 111/4  & \textbf{91.4\%} & \textbf{64.3\%}$^\dagger$ & 10.0\% & Config-spec. \\
 & exploration ratio  & 119 & $+0.018$ &  61/44 & \textbf{89.4\%} & 34.7\%                    & 16.1\% & Config-spec. \\
 & modification ratio & 119 & $+0.026$ &  60/42 & \textbf{88.7\%} & 36.3\%                    & 14.3\% & Config-spec. \\
 & repetition rate    & 119 & $+0.066$ &  86/15 & \textbf{81.4\%} & 39.1\%                    & 13.1\% & Config-spec. \\
 & test ratio         & 119 & $-0.036$ &  29/55 & \textbf{75.7\%} & 34.2\%                    & 32.1\% & Config-spec. \\
 & error rate         & 119 & $+0.006$ &  47/48 & 69.4\%          & 35.5\%                    & 25.9\% & Moderate \\
 & cascade rate       & 119 & $+0.046$ &  76/13 & 45.0\%          & \textbf{44.4\%}$^\dagger$ & 21.3\% & Moderate \\
\addlinespace
\multicolumn{9}{l}{\textit{CFG features}} \\
 & motif entropy       & 119 & $+0.161$ & 100/13 & \textbf{91.7\%} & 48.0\%$^\dagger$          & 16.7\% & Config-spec. \\
 & transition entropy  & 119 & $+0.218$ & 106/9  & \textbf{91.5\%} & \textbf{51.4\%}$^\dagger$ & 17.5\% & Config-spec. \\
 & revisit rate        & 119 & $+0.263$ & 113/3  & \textbf{90.3\%} & \textbf{50.8\%}$^\dagger$ & 20.3\% & Config-spec. \\
 & backtrack rate      & 119 & $+0.154$ & 102/8  & \textbf{88.9\%} & 43.8\%$^\dagger$          & 20.7\% & Config-spec. \\
 & self-loop rate      & 119 & $+0.111$ &  95/18 & \textbf{85.9\%} & 46.1\%$^\dagger$          & 18.3\% & Config-spec. \\
 & post-error motif    & 119 & $-0.016$ &  37/63 & 72.9\%          & 39.1\%                    & 27.3\% & Moderate \\
\bottomrule
\end{tabular}
\begin{tablenotes}
\footnotesize
\item $K$: valid configurations. $\bar r$: inverse-variance-weighted mean of rank-biserial $r$ (positive = lower feature value associated with resolution). $n_+/n_-$: count of configurations with positive vs.\ negative per-configuration $r$ (residual are zero-effect). $R^2_{\text{FW}}$ ($R^2_{\text{LLM}}$): fraction of between-configuration variance explained by framework identity (LLM family) in a random-effects meta-regression. Classification: universal ($I^2 < 25\%$), moderate ($25$--$75\%$), configuration-specific ($\geq 75\%$). $\dagger$ marks $R^2_{\text{FW}}$ values that exceed the framework permutation null at $\alpha = 0.05$ (\S\ref{sec:robustness}).
\end{tablenotes}
\end{threeparttable}
\end{table*}

\paragraph{Ten of 13 features are configuration-specific.}
The configuration-specific group spans both the action side (mean turns, exploration ratio, modification ratio, repetition rate, test ratio) and the CFG side (motif entropy, transition entropy, revisit rate, backtrack rate, self-loop rate). Cascade rate ($I^2 = 45.0\%$), error rate ($69.4\%$), and post-error motif ratio ($72.9\%$) fall in the moderate band. No feature approaches universality; even the most stable predictors show substantial between-configuration variation in effect magnitude.

\begin{figure*}[!htb]
\centering
\includegraphics[width=0.85\linewidth]{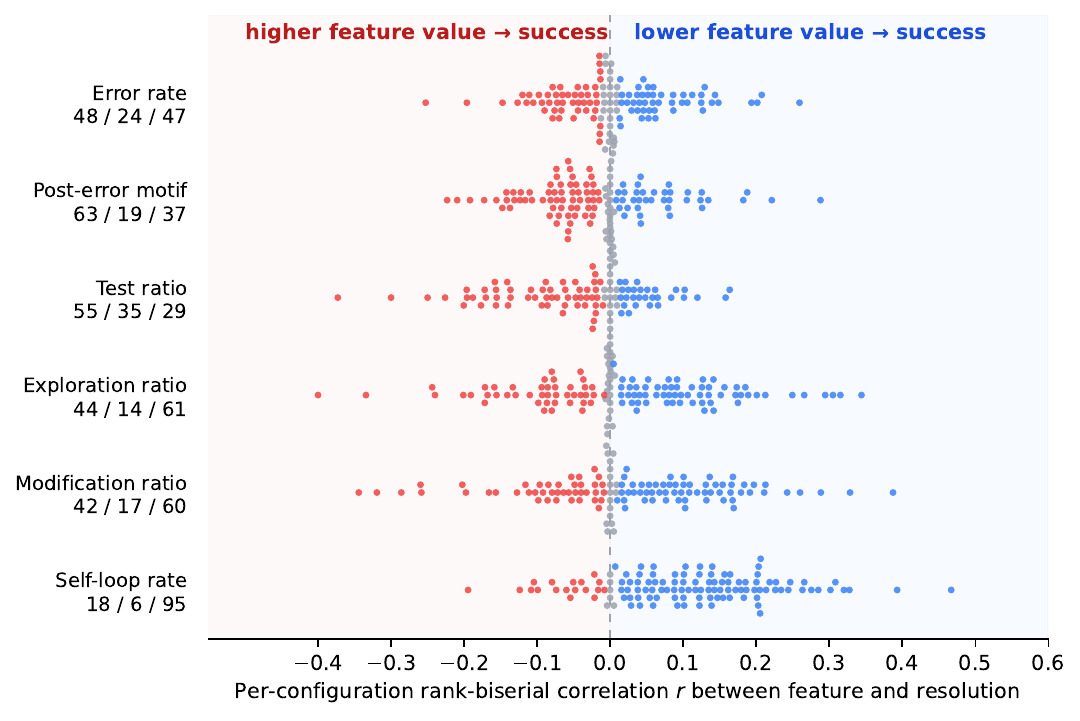}
\caption{Per-configuration effects of six direction-unstable continuous features on resolution. Each row plots 119 configurations as a horizontal beeswarm of rank-biserial $r$; red = higher feature value associated with success, blue = lower, grey = no measurable direction. Split labels (red / grey / blue) match the $n_+/n_-$ entries in Table~\ref{tab:behavior_outcome}. Sign convention: positive $r$ = lower feature value associated with resolution.}
\label{fig:direction_unstable}
\end{figure*}

\paragraph{Directional disagreement is real and substantial.}
The $n_+/n_-$ columns reveal that for six features, a non-trivial fraction of configurations show the opposite direction from the weighted mean: error rate ($47/48$), post-error motif ($37/63$), test ratio ($29/55$), exploration ratio ($61/44$), modification ratio ($60/42$), and self-loop rate ($95/18$). For error rate and post-error motif, the minority share approaches parity: the same observable signal predicts resolution in roughly half the configurations and failure in the other half. Figure~\ref{fig:direction_unstable} plots the full per-configuration distribution for these six features; each row arranges 119 configurations as a beeswarm of rank-biserial $r$, with red dots marking ``higher feature value associates with success'' and blue dots marking the opposite direction.

\paragraph{Framework identity is the stronger moderator for trajectory-shape features.}
Framework identity numerically exceeds LLM family for all 13 features, but the gap is statistically meaningful only for a subset. Mean turns is the strongest case at $6.4\times$ ($R^2_{\text{FW}} = 64.3\%$ versus $R^2_{\text{LLM}} = 10.0\%$); six other trajectory-shape features (Table~\ref{tab:behavior_outcome}, $\dagger$) cluster at $2$--$3\times$ and clear the permutation chance baseline (\S\ref{sec:robustness}). For the remaining six features, the observed $R^2_{\text{FW}}$ falls within the chance baseline; cross-configuration heterogeneity is real ($I^2 = 69$--$89\%$ in this group) but neither moderator separates from noise.

\paragraph{Connection to RQ1.}
RQ1 showed that framework identity shapes \textit{behavior} ($\eta^2 = 0.17$--$0.90$). RQ3 now shows that framework identity also shapes the \textit{behavior--outcome relationship} ($R^2_{\text{FW}} = 34$--$64\%$). The two results close a loop: framework $\to$ behavior, framework $\to$ (behavior $\to$ outcome), which together suggest that behavioral guidance should be framework-aware rather than extracted from a single reference framework and exported.

\begin{rqbox}[Answer to RQ3]
The predictive value of behavioral features is configuration-dependent: 10 of 13 continuous features are configuration-specific ($I^2 = 75.7\%$--$91.7\%$); three are moderate; no feature approaches universality. Framework identity is the stronger moderator for the seven trajectory-shape features that clear the permutation chance baseline (Table~\ref{tab:behavior_outcome}, $\dagger$), with mean turns as the largest case. For the remaining six features, neither moderator separates from the chance baseline (\S\ref{sec:robustness}). Six features show substantial directional disagreement across configurations: the same observable signal carries opposite meaning for different agents even without any type-level grouping.
\end{rqbox}

\FloatBarrier

\subsection{Robustness Checks}
\label{sec:robustness}

The framework $R^2$ values in Tables~\ref{tab:patterns} and~\ref{tab:behavior_outcome} are observed point estimates from a 43-level dummy regression. To check that the framework moderator separates from sampling noise, we compute three diagnostics: (i)~95\% bootstrap percentile intervals on $R^2$ (2{,}000 resamples of configurations), (ii)~a one-sided permutation null built by shuffling framework labels among configurations and recomputing $R^2$ (2{,}000 shuffles), and (iii)~exhaustive leave-one-framework-out resampling. The permutation $p$-values reported below test label structure at the configuration level ($K \leq 119$) rather than pooled-effect non-zeroness; the per-trajectory sample-size critique that motivates our usual avoidance of $p$-values (\S\ref{sec:per_config_meta}) therefore does not apply.

The permutation null is informative because the 43-level dummy regression overfits even random labels: across the 13 continuous features the null mean is $0.33$ with 95th percentile $0.45$, so an observed $R^2_{\text{FW}}$ below this 95th percentile cannot be distinguished from a chance label assignment. Seven of 13 continuous features clear the threshold at $\alpha = 0.05$ (Table~\ref{tab:behavior_outcome}, $\dagger$). Mean turns is the strongest case ($R^2_{\text{FW}} = 0.64$, 95\% CI $[0.52, 0.81]$, permutation $p < 0.001$, leave-one-framework-out range $[0.53, 0.80]$); the other six (five control-flow topology features and cascade rate) follow at $R^2_{\text{FW}} = 0.45$--$0.51$ with permutation $p \in [0.003, 0.034]$. The remaining six features fall within the chance baseline ($R^2_{\text{FW}} = 0.34$--$0.40$, permutation $p \in [0.17, 0.41]$). Among the seven binary patterns in RQ2, only P6 (low late entropy) clears the threshold ($R^2_{\text{FW}} = 0.49$, permutation $p = 0.016$); for the remaining six, binary discretization saturates the contingency table ($n_+/n_-$ heavily skewed in five of seven), which limits meta-regression power.

For four signals (P4 fast cascade recovery, P7 test-after-modify, test ratio, and post-error motif ratio), the LLM-family moderator separately exceeds the chance baseline while the framework moderator does not, indicating that the LLM is the stronger explanatory factor on these signals. The cross-configuration heterogeneity reported in \S\ref{sec:rq2} and~\S\ref{sec:rq3} ($I^2$ values, direction splits) is unaffected by these diagnostics, which constrain only the question of which layer drives the variance. Tables~\ref{tab:patterns} and~\ref{tab:behavior_outcome} mark the rows that pass the permutation test with $\dagger$; the full per-feature robustness table (bootstrap CIs, permutation $p$ for both moderators, leave-one-framework-out ranges) is available as supplementary material.

%% file: sections/discussion.tex
\section{Discussion}
\label{sec:discussion}

The central finding is a two-part claim about the 126-configuration ecosystem. First, the same observable behavioral signal (e.g., error rate, post-error motif ratio, test ratio, exploration ratio) carries opposite meaning for a substantial fraction of configurations, with sign disagreement visible at the configuration level (\S\ref{sec:insight1}). Second, when we ask \textit{which layer of the agent stack} explains that heterogeneity, framework identity is the stronger moderator for trajectory-shape features (\S\ref{sec:robustness}); for action-composition and raw error features, neither moderator separates from chance. These two facts together explain why heuristics exported from one framework routinely fail on another: where a moderator can be pinned down, the framework reshapes what the signal means, while other divergences remain observed but not yet attributable.

\subsection{Same Signal, Different Semantics}
\label{sec:insight1}

The per-configuration meta-analysis shows direction disagreement, not merely effect-magnitude variation. Across the 13 continuous features in RQ3, six features show direction splits across configurations: five with at least one-third of configurations on each side (Table~\ref{tab:signal_semantics}), and self-loop rate at 95/18; three of the seven binary patterns in RQ2 show the same pattern. Error rate is the most balanced example: 47 configurations show lower error rate associated with resolution (cleaner execution $\to$ success), while 48 show the opposite (higher error rate $\to$ success, plausibly indicating tolerated iterative revision).

\begin{table*}[!htb]
\caption{Same Signal, Different Semantics: Configuration-Level Direction Splits}
\label{tab:signal_semantics}
\centering
\small
\begin{tabularx}{\linewidth}{l>{\centering\arraybackslash}X>{\centering\arraybackslash}X>{\raggedright\arraybackslash}X}
\toprule
\textbf{Signal} & \textbf{Lower-value $\to$ success ($n_+$)} & \textbf{Higher-value $\to$ success ($n_-$)} & \textbf{Weighted mean} \\
\midrule
\multicolumn{4}{l}{\textit{Continuous features (RQ3)}} \\
Error rate                & 47 & 48 & $\bar r = +0.006$ (near null) \\
Post-error motif ratio    & 37 & 63 & $\bar r = -0.016$ (leans higher $\to$ success) \\
Test ratio                & 29 & 55 & $\bar r = -0.036$ (leans higher $\to$ success) \\
Exploration ratio         & 61 & 44 & $\bar r = +0.018$ (leans lower $\to$ success) \\
Modification ratio        & 60 & 42 & $\bar r = +0.026$ (leans lower $\to$ success) \\
\addlinespace
\multicolumn{4}{l}{\textit{Binary patterns (RQ2), $n_+/n_-$ on pattern presence $\to$ resolution}} \\
P4 Fast cascade recovery  & 44 & 54 & $\bar V = +0.001$ (near null) \\
P6 Low late entropy       & 45 & 46 & $\bar V = -0.008$ (near null) \\
P7 Test-after-modify      & 50 & 33 & $\bar V = +0.017$ (weakly positive overall) \\
\bottomrule
\end{tabularx}
\end{table*}

We interpret this divergence through the universal problem-solving workflow SE agents share: \textbf{locate $\to$ modify $\to$ verify}. Every agent must find the relevant code, change it, and verify the change; different frameworks execute the three stages with different degrees of agent autonomy, iteration budget, and failure tolerance, which changes what each behavioral signal means in context. In long-deliberate workflows, low error rates serve as evidence of clean localization and correlate with success; in tight modify-verify loops, elevated error rates reflect committed revision and correlate with success. The observable action is identical; the surrounding workflow determines whether it signals discipline or collapse.

\subsection{Two Classes of Behavioral Guidance}
\label{sec:insight2}

The per-configuration evidence separates behavioral signals into two transferability classes (Table~\ref{tab:guidelines}). Direction-stable signals (shorter trajectories, fewer revisits, lower motif and transition entropy, and lower backtrack rate) transfer as qualitative principles, with $\geq 88\%$ of configurations agreeing on the sign of the effect.\footnote{The $88\%$ threshold includes motif entropy ($88.5\%$ positive); a stricter $\geq 90\%$ threshold reclassifies motif entropy as direction-unstable, but the broader two-class structure is invariant.} These six measurements are correlated and effectively form a single construct of trajectory complexity, so the direction-stable result is most accurately read as ``trajectory complexity transfers as a qualitative principle''. Even within this class, magnitude varies sharply: per-configuration effects span $-0.38$ to $+0.58$ at $I^2 = 86\%$--$92\%$, and framework identity explains $2.1\times$--$6.4\times$ more of the magnitude variation than LLM family does ($R^2_{\text{FW}} = 32\%$--$64\%$ versus $R^2_{\text{LLM}} = 10\%$--$21\%$), so a numeric target (e.g., a turn-count cap) requires per-framework recalibration. Direction-unstable signals carry no transferable rule. The class comprises error rate (47/48 split), test ratio (29/55), post-error motif (37/63), exploration ratio (61/44), modification ratio (60/42), self-loop rate (95/18), and binary patterns P4, P6, P7; an exported direction-unstable rule misleads a substantial fraction of adopters. The minimum condition for a behavioral finding to translate across frameworks is reporting the framework alongside the result, as context rather than caveat.

\begin{table*}[!htb]
\caption{Two Classes of Behavioral Guidance, by Per-Configuration Evidence}
\label{tab:guidelines}
\centering
\small
\begin{tabularx}{\linewidth}{>{\raggedright\arraybackslash}p{2.4cm}X>{\raggedright\arraybackslash}p{2.6cm}}
\toprule
\textbf{Class} & \textbf{Representative signals and evidence} & \textbf{Transfer} \\
\midrule
Direction-stable & Shorter trajectories (P5, 93/8), fewer revisits (113/3), lower motif entropy (100/13), lower transition entropy (106/9), lower backtrack rate (102/8), shorter mean turns (111/4); $\geq\!88\%$ of configurations agree on direction. Per-configuration effect sizes span $-0.38$ to $+0.58$ at $I^2 = 86\%$--$92\%$; framework identity explains $2.1\times$--$6.4\times$ more of the magnitude variation than LLM family ($R^2_{\text{FW}} = 32\%$--$64\%$ vs.\ $R^2_{\text{LLM}} = 10\%$--$21\%$). & Principle transfers; numeric targets require framework-specific calibration \\
\addlinespace[2pt]
Direction-unstable & Error rate (47/48), post-error motif (37/63), test ratio (29/55), exploration ratio (61/44), modification ratio (60/42), self-loop rate (95/18); P4 fast recovery (44/54), P6 low late entropy (45/46), P7 test-after-modify (50/33). Substantial fraction of configurations show opposite sign. & No universal transfer \\
\bottomrule
\end{tabularx}
\end{table*}

\subsection{Framework as the Primary Lever}
\label{sec:practitioners}

When an SE agent underperforms, practitioners face a binary choice: upgrade to a stronger LLM, or redesign the framework while keeping a cost-effective model. The within-type controlled comparison (Table~\ref{tab:dir_ab}) shows these strategies are not equivalent, and the appropriate choice depends on trajectory type (Table~\ref{tab:strategy_guide}).

Type~3 (49.8\% resolution) and Type~5 (46.3\%) are strongly framework-dominated (FW wins 94\% and 83\%; median FW $\eta^2 = 0.48$ and $0.36$ vs.\ LLM $\eta^2 = 0.11$ and $0.02$): the primary bottleneck is framework design, and framework redesign is the higher-leverage investment. Type~1 (42.1\%, long-exploratory) reverses the pattern with LLM dominance (FW wins only 33\%, median LLM $\eta^2 = 0.28$ vs.\ FW $\eta^2 = 0.08$): the framework infrastructure already generates long deliberate trajectories, so the incremental bottleneck is reasoning quality and stronger models add more value. Type~2 (57.6\%, framework-leaning at 67\%; median FW $\eta^2 = 0.22$ vs.\ LLM $\eta^2 = 0.02$) already achieves the highest resolution rate, where moderate gains may come from either lever. Type~4 lacks sufficient same-type tracer pairs for a direct comparison.

To apply these recommendations, a practitioner first identifies which trajectory type their agent produces by computing the behavioral features from Table~\ref{tab:features} on a sample of its trajectories, then prioritizes the type-indicated lever. The transferability of any specific behavioral rule should be checked against Table~\ref{tab:guidelines}: direction-stable principles apply broadly, while direction-unstable rules (e.g., test-after-modify guidance derived from OpenHands studies) may not transfer to a Type~1 framework and can harm performance if applied uncritically.

\begin{table*}[!htb]
\caption{Type-Level Descriptive Summary and Improvement Lever (from RQ1 within-type FW vs.\ LLM comparison; Frm = framework-dominated, Mdl = LLM-dominated; --- = insufficient data for Type~4)}
\label{tab:strategy_guide}
\centering
\small
\begin{tabular}{lccll}
\toprule
\textbf{Type} & \textbf{Res.} & \textbf{T-Ent.} & \textbf{Behavioral marker} & \textbf{Lev.} \\
\midrule
Type~1  & 42.1\% & 1.25 & Long ($\approx\!44$ turns), low front-loading & Mdl \\
Type~2  & 57.6\% & 1.16 & Moderate, partial front-loading               & Frm \\
Type~3  & 49.8\% & 1.48 & Moderate, highest transition entropy          & Frm \\
Type~4  & 42.1\% & 0.29 & Compact ($\approx\!10$ turns), lowest entropy & --- \\
Type~5  & 46.3\% & 0.50 & Compact ($\approx\!14$), front-loading 0.96   & Frm \\
\bottomrule
\end{tabular}
\end{table*}

\subsection{Implications}
\label{sec:implications}

\paragraph{For framework designers.}
Three guidelines follow from the per-configuration evidence.

\textit{(1) Build framework-aware recovery.} Post-error motif ratio is direction-unstable across configurations: higher post-error motif associates with resolution in 63 configurations (persistent repair pays off) but with failure in 37 (spiraling into an unrecoverable state). On this signal it is the LLM rather than the framework that emerges as the stronger moderator under permutation testing (\S\ref{sec:robustness}), so retry limits and re-localization triggers should be calibrated against the target configuration's own trajectory history, with the calibration tested separately for each LLM in use.

\textit{(2) Do not enforce direction-unstable signals as universal rules.} Exploration budgets, test-after-modify, and error-rate thresholds fall in the direction-unstable class (Table~\ref{tab:guidelines}) and should not be hard-wired into system prompts or scoring functions reused across frameworks.

\textit{(3) Expose behavioral telemetry.} Action sequence patterns, exploration ratio, cascade length, phase transition timing, and CFG revisit and backtrack rates can serve as real-time diagnostic signals. Backtrack rate is the most directionally consistent warning across the ecosystem (102 of 110 non-zero configurations show elevated backtracking associated with worse outcomes), making it the strongest framework-agnostic monitoring metric. Frameworks that expose these metrics let practitioners fit configuration-specific thresholds and intervene before failures become irrecoverable.

\paragraph{For researchers.}
Framework identity should be treated as a primary independent variable rather than a control. Across the 20 behavioral signals tested, nine show substantial per-configuration sign disagreement (six of 13 continuous features, three of seven binary patterns; Table~\ref{tab:guidelines}); for 10 of 13 continuous features, framework identity explains at least twice the between-configuration variance that LLM family does. Patterns that appear fundamental within one framework can be artifacts of its workflow, tool surface, or system-prompt conventions. The methodological recommendation is to run a per-configuration meta-analysis with framework identity and LLM family as moderators: $I^2$ quantifies how much of a finding transfers across configurations, and $R^2$ decomposition identifies which layer of the stack drives any non-transfer. Most prior behavioral studies focus on one framework or a few structurally similar variants; their findings are valid scope-limited claims rather than universal ones. Reporting the framework, and where available the per-configuration effect distribution, alongside behavioral findings should be a first-class characterization of study scope, not a footnote.

\subsection{Future Work}
\label{sec:future}

\paragraph{Framework-aware intervention studies.}
A small-scale mechanism probe (14 paired SWE-bench tasks on SWE-Agent, same task, same LLM, with and without error-triggered re-localization) improves resolution from 2/14 to 3/14, showing the post-error signal is intervention-sensitive within SWE-Agent. The $n = 14$ probe is directionally consistent but underpowered; a properly powered study (hundreds of paired tasks per framework) would test whether targeted changes (post-error re-localization, structured patching limits, exploration budgets calibrated to the target framework's distribution) shift behavior in the predicted direction across the direction splits in Table~\ref{tab:signal_semantics}.

\paragraph{Cross-benchmark validation.}
Replicating the per-configuration meta-analysis on tasks beyond SWE-bench (feature development, refactoring, cross-language repositories) would test whether the framework-dominated heterogeneity pattern is specific to the bug-fix setting or a general property of SE agent ecosystems.

\paragraph{Finer-grained moderators.}
Framework identity is a bundle treatment that mixes tool surface, system prompt, iteration budget, and context management. Decomposing it into distinct moderator variables, each tested separately in the meta-regression, could support more targeted design recommendations without growing the taxonomy.

\paragraph{Real-time behavioral monitoring.}
The metrics identified here (cascade length, exploration ratio, phase transition timing) could be implemented as real-time monitoring signals in production deployments. Thresholds fit against the target configuration's own historical distribution, rather than exported defaults, would enable automated intervention when failure-associated patterns are detected.

\paragraph{Identifying components within the framework bundle.}
A controlled ablation that varies one framework component at a time (system prompt, tool surface, iteration budget, context-management strategy) on the same architecture would complement the observational evidence here and identify which sub-component drives the framework effect on each behavioral signal.

%% file: sections/threats.tex
\section{Threats to Validity}
\label{sec:threats}

\paragraph{Internal validity.} All analyses are observational; associations do not establish causality. Each of the 126 configurations resolves a partly different subset of SWE-bench, so per-configuration effect sizes mix behavioral semantics with per-configuration task-difficulty differences. A task-difficulty alternative therefore remains compatible with the observed direction splits: some configurations may be easy enough that feature $f$ correlates one way, while harder configurations reverse the sign. Our claims are accordingly about covariation with framework identity, not causal attribution to specific design components.

\paragraph{Permutation null and overfitting baseline.} The meta-regression with 43 framework dummies has a non-trivial overfitting baseline: random framework labels yield $R^2$ averaging $0.33$ across the 13 continuous features, with a 95th percentile of $0.45$. We evaluate moderator strength against this baseline rather than against zero, and report per-feature outcomes in \S\ref{sec:robustness}. For features below the baseline, the cross-configuration heterogeneity is real ($I^2$ values and direction splits in \S\ref{sec:rq2}, \S\ref{sec:rq3}) but is not cleanly attributable to either moderator.

\paragraph{Prompt engineering confound.} ``Framework'' in the multi-tracer design is a bundle treatment: frameworks differ in interaction architecture but also in system prompt content, tool descriptions, instruction format, and context length, and trajectory data alone cannot separate these components. RQ1's framework effect therefore encompasses both architectural and prompt-design variation.

\paragraph{Sparse moderator matrix.} The (framework, LLM family) cross has only $9.3\%$ cell coverage (74 of 792); 36 of 44 frameworks span only a single LLM family. The framework and LLM-family moderators are therefore not independently identified from the joint matrix, and the framework-vs-LLM contrast in the meta-regression should be read alongside the controlled bash-only baseline and the three-tracer natural experiment in \S\ref{sec:design}, which hold one layer fixed where the data support it. The full coverage matrix is provided as supplementary material.

\paragraph{Framework version confound.} Within the bash-only subset, mini-swe-agent evolved across versions v0.0.0 through v1.17.2. Some behavioral differences attributed to LLM identity may reflect framework improvements co-occurring with newer model releases. We mitigate this by using mini-swe-agent v1.0.0 (9 models, 6 LLM families) as the primary controlled baseline; the multi-tracer design uses exact version matches within each tracer.

\paragraph{Threshold sensitivity.} Binary pattern thresholds (P1--P7) are derived from the bash-only subset. Seven alternative threshold settings recompute the per-configuration meta-analysis under varied trajectory-length, cascade, and entropy thresholds. Four patterns (P1, P2, P4, P7) are stable across all seven settings. P5 (shorter trajectory) remains classified as configuration-specific in all seven settings ($I^2 = 82$--$90\%$). P3 (shorter cascades) shifts from $50.8\%$ to $56.8\%$ under a looser cascade threshold but remains moderate. P6 (low late entropy) is the most threshold-sensitive: its $I^2$ ranges from $54.2\%$ to $80.0\%$ depending on the entropy median, crossing the $75\%$ boundary in one setting. P7 sits at the moderate/configuration-specific boundary ($I^2 = 75.0\%$) in every setting.

\paragraph{Construct validity.} All trajectories were generated under the 100\% oracle setting (full test visibility), which may inflate test-related signals relative to deployment conditions. In particular, \texttt{test\_ratio} and the test-after-modify heuristic (P7) are confounded by the oracle: agents may strategically reduce testing because the oracle already reveals pass/fail status, making observed test ratios partly an artifact of the evaluation protocol. Our findings characterize behavior under maximum information availability and should not be generalized to other oracle settings without further study. The six-category action taxonomy abstracts over implementation differences (a fault-localization API and a raw \texttt{grep} both map to Exploration), enabling cross-framework comparison at the cost of within-framework semantic precision. The single configuration with $>$50\% unknown actions (a tool-call agent at $65\%$) is excluded from action-distribution analyses.

\paragraph{Tool-call classifier scope.} Beyond the single $65\%$-unknown outlier above, tool-call agents (whose actions are API calls rather than bash commands) produce elevated unknown rates under the regex-based classifier. We exclude affected configurations from action-distribution analyses but retain them for resolution-rate comparisons; the net effect is a $\approx 1.6\%$ reduction in usable trajectories for action-distribution features, concentrated in Type~3 and Type~4.

\paragraph{Taxonomy boundaries.} The five types impose discrete cuts on a continuous behavioral space and are used only for descriptive reporting (the cluster-count choice and silhouette comparison are documented in \S\ref{sec:design}, ``Cluster count''). The same framework appears in multiple types across LLM configurations, reflecting genuine behavioral variation rather than classification instability.

\paragraph{External validity.} All trajectories target SWE-bench Verified (500 Python repository issues). Results may not generalize to other languages, task types (feature development, refactoring), or newer benchmarks such as SWE-Bench-Pro. Our claims are therefore about a large historical and ecosystem-level slice of publicly available SE-agent behavior, not about permanent benchmark-specific leaderboard rankings. Type~4 and Type~5 are compact-trajectory types (median 10 and 14 turns respectively); their behavioral findings reflect the structural constraints of short interaction budgets and should not be read as upper bounds on agent capability in less constrained settings.

\paragraph{Within-type heterogeneity.} Type~3 spans 48 configurations with resolution rates ranging from near zero (five configurations producing zero resolved tasks under weaker LLMs) up to $0.77$. Even after excluding bottom-tier outliers, the 10th--90th percentile interval is $0.06$--$0.74$ ($\approx$$12\times$ spread). RQ3 rank-biserial values for this group are weighted averages over this diverse population and should not be interpreted as representative of any individual framework. Type~1 (24 configurations) exhibits similar internal variation.

\paragraph{Reliability.} Each feature aggregates tens of per-configuration effect sizes (every continuous feature pools 119 of the 126 configurations; the seven binary patterns pool between 40 and 110, with the exact per-feature $K$ reported in Tables~\ref{tab:patterns} and~\ref{tab:behavior_outcome}). The directional-disagreement evidence ($n_+/n_-$) is a raw configuration count independent of $I^2$ precision. Four cascade definitions (cascade min length $\in \{2,3,4,5\}$) leave P3 stable at $I^2 \approx 50.8\%$ and the continuous cascade-rate feature in the moderate band; threshold sensitivity is reported above. The Direction-stable versus Direction-unstable classification in \S\ref{sec:insight2} depends on the chosen agreement threshold ($\geq 88\%$); borderline signals near this boundary, notably motif entropy at $88.5\%$, reclassify under stricter thresholds, but the broader two-class structure is invariant.

%% file: sections/conclusion.tex
\section{Conclusion}
\label{sec:conclusion}

The single-configuration behavioral study has a structural blind spot: it cannot detect when a pattern's predictive meaning flips across the ecosystem. Applying a per-configuration random-effects meta-analysis to 64{,}380 trajectories from 126 $\langle$framework, LLM$\rangle$ configurations reveals that this is the common case, not the exception. Of 13 continuous behavioral features, 10 are configuration-specific ($I^2 \geq 75\%$), and six exhibit substantial directional disagreement (error rate at 47/48, post-error motif at 37/63, test ratio at 29/55, and three additional features). Of seven binary patterns drawn from prior single-framework studies, two are configuration-specific while two more reverse direction. The same observable signal can be evidence of discipline in one workflow and collapse in another, depending on how the framework structures the locate-modify-verify cycle (\S\ref{sec:insight1}).

Framework identity is the stronger moderator for trajectory-shape features, with mean turns as the largest case ($R^2_{\text{FW}} = 64\%$ versus $R^2_{\text{LLM}} = 10\%$). RQ1's multi-tracer analysis confirms framework dominance at the trajectory level (framework $\eta^2 = 0.17$--$0.90$ when the LLM is held fixed); for action-composition features and raw error counters, neither moderator separates from chance (\S\ref{sec:robustness}). Behavioral signals therefore split into two transferability classes (Table~\ref{tab:guidelines}): direction-stable signals (shorter trajectories, fewer revisits, lower entropy) carry qualitative principles across frameworks but require per-framework recalibration of magnitude; direction-unstable signals carry no transferable rule.

The practical implication is immediate: before applying a behavioral rule, verify that its direction holds on the target framework, not merely that it appeared in a prior single-framework study (\S\ref{sec:results}). The research implication is structural: framework identity is a primary independent variable, not a control variable, and the per-configuration meta-analysis we apply is a route to making the scope of behavioral findings explicit. Conclusions drawn within one framework are scope-limited claims, not universal rules, until validated across the ecosystem.

\section*{Data and Code Availability}
Trajectories are publicly available from the SWE-bench Experiments repository~\cite{jimenez2024swebench} and the Docent platform (both accessed January~2026). The analysis code, processed datasets, per-configuration effect-size tables, and a packaged analysis toolkit that re-runs the entire per-configuration meta-analysis pipeline are released at \url{https://github.com/Marvinmw/agent-design-artifact}. The toolkit accepts trajectory inputs in the formats used by the 43 frameworks studied here and produces $I^2$, direction-split, and $R^2_{\text{FW}}/R^2_{\text{LLM}}$ diagnostics for arbitrary behavioral features, so that other researchers can apply the same cross-configuration validity test to new frameworks or new behavioral patterns.